\documentclass[twocolumn]{aastex631}

\usepackage{amsmath}
\usepackage{verbatim}
\usepackage{hyperref}

\newcommand{\onenineteen}{KMT-2021-BLG-0119}
\newcommand{\oneninetwo}{KMT-2021-BLG-0192}
\newcommand{\twotwoninefour}{KMT-2021-BLG-2294}

\newcommand{\pivec}{\mbox{\boldmath $\pi$}}

\newcommand{\thetavec}{\mbox{\boldmath $\theta$}}
\newcommand{\pieN}{\mbox{$\pi_{{\rm E},N}$}}
\newcommand{\pieE}{\mbox{$\pi_{{\rm E},E}$}}

\lefthead{} 
\righthead{}

\begin{document}

\title{Mass Production of 2021 KMTNet Microlensing Planets III: Analysis of Three Giant Planets
}

\author{In-Gu Shin} 
\affiliation{Center for Astrophysics $|$ Harvard \& Smithsonian 60 Garden St., Cambridge, MA 02138, USA}
\author{Jennifer C. Yee}
\affiliation{Center for Astrophysics $|$ Harvard \& Smithsonian 60 Garden St., Cambridge, MA 02138, USA}
\author{Andrew Gould}
\affiliation{Max Planck Institute for Astronomy, K\"onigstuhl 17, D-69117 Heidelberg, Germany}
\affiliation{Department of Astronomy, The Ohio State University, 140 W. 18th Ave., Columbus, OH 43210, USA}
\author{Kyu-Ha Hwang}
\affiliation{Korea Astronomy and Space Science Institute, Daejon 34055, Republic of Korea}
\author{Hongjing Yang}
\affiliation{Department of Astronomy and Tsinghua Centre for Astrophysics, Tsinghua University, Beijing 100084, China}
\author{Ian A. Bond}
\affiliation{Institute of Natural and Mathematical Sciences, Massey University, Auckland 0745, New Zealand}
\collaboration{7}{(Leading authors),}
%
\author{Michael D. Albrow} 
\affiliation{University of Canterbury, Department of Physics and Astronomy, Private Bag 4800, Christchurch 8020, New Zealand}
\author{Sun-Ju Chung}
\affiliation{Korea Astronomy and Space Science Institute, Daejon 34055, Republic of Korea}
\author{Cheongho Han}
\affiliation{Department of Physics, Chungbuk National University, Cheongju 28644, Republic of Korea}
\author{Youn Kil Jung}
\affiliation{Korea Astronomy and Space Science Institute, Daejon 34055, Republic of Korea}
\affiliation{University of Science and Technology, Korea, (UST), 217 Gajeong-ro, Yuseong-gu, Daejeon 34113, Republic of Korea}
\author{Yoon-Hyun Ryu}
\affiliation{Korea Astronomy and Space Science Institute, Daejon 34055, Republic of Korea}
\author{Yossi Shvartzvald}
\affiliation{Department of Particle Physics and Astrophysics, Weizmann Institute of Science, Rehovot 76100, Israel}
\author{Weicheng Zang}
\affiliation{Department of Astronomy and Tsinghua Centre for Astrophysics, Tsinghua University, Beijing 100084, China}
%
\author{Sang-Mok Cha}
\affiliation{Korea Astronomy and Space Science Institute, Daejon 34055, Republic of Korea}
\affiliation{School of Space Research, Kyung Hee University, Yongin, Kyeonggi 17104, Republic of Korea}
\author{Dong-Jin Kim}
\affiliation{Korea Astronomy and Space Science Institute, Daejon 34055, Republic of Korea}
\author{Seung-Lee Kim} 
\affiliation{Korea Astronomy and Space Science Institute, Daejon 34055, Republic of Korea}
\author{Chung-Uk Lee}
\affiliation{Korea Astronomy and Space Science Institute, Daejon 34055, Republic of Korea}
\author{Dong-Joo Lee}
\affiliation{Korea Astronomy and Space Science Institute, Daejon 34055, Republic of Korea}
\author{Yongseok Lee}
\affiliation{Korea Astronomy and Space Science Institute, Daejon 34055, Republic of Korea}
\affiliation{School of Space Research, Kyung Hee University, Yongin, Kyeonggi 17104, Republic of Korea}
\author{Byeong-Gon Park}
\affiliation{Korea Astronomy and Space Science Institute, Daejon 34055, Republic of Korea}
\affiliation{University of Science and Technology, Korea, (UST), 217 Gajeong-ro, Yuseong-gu, Daejeon 34113, Republic of Korea}
\author{Richard W. Pogge}
\affiliation{Department of Astronomy, The Ohio State University, 140 W. 18th Ave., Columbus, OH 43210, USA}
\collaboration{16}{(The KMTNet Collaboration),}
\author{Fumio Abe}
\affiliation{Institute for Space-Earth Environmental Research, Nagoya University, Nagoya 464-8601, Japan}
\author{Richard Barry}
\affiliation{Code 667, NASA Goddard Space Flight Center, Greenbelt, MD 20771, USA}
\author{David P.~Bennett}
\affiliation{Code 667, NASA Goddard Space Flight Center, Greenbelt, MD 20771, USA}
\affiliation{Department of Astronomy, University of Maryland, College Park, MD 20742, USA}
\author{Aparna Bhattacharya}
\affiliation{Code 667, NASA Goddard Space Flight Center, Greenbelt, MD 20771, USA}
\affiliation{Department of Astronomy, University of Maryland, College Park, MD 20742, USA}
\author{Hirosane Fujii}
\affiliation{Institute for Space-Earth Environmental Research, Nagoya University, Nagoya 464-8601, Japan}
\author{Akihiko Fukui}
\affiliation{Department of Earth and Planetary Science, Graduate School of Science, The University of Tokyo, 7-3-1 Hongo, Bunkyo-ku, Tokyo 113-0033, Japan}
\affiliation{Instituto de Astrof\'isica de Canarias, V\'ia L\'actea s/n, E-38205 La Laguna, Tenerife, Spain}
\author{Yuki Hirao}
\affiliation{Department of Earth and Space Science, Graduate School of Science, Osaka University, Toyonaka, Osaka 560-0043, Japan}
\author{Stela Ishitani Silva}
\affiliation{Department of Physics, The Catholic University of America, Washington, DC 20064, USA}
\affiliation{Code 667, NASA Goddard Space Flight Center, Greenbelt, MD 20771, USA}
\author{Yoshitaka Itow}
\affiliation{Institute for Space-Earth Environmental Research, Nagoya University, Nagoya 464-8601, Japan}
\author{Rintaro Kirikawa}
\affiliation{Department of Earth and Space Science, Graduate School of Science, Osaka University, Toyonaka, Osaka 560-0043, Japan}
\author{Iona Kondo}
\affiliation{Department of Earth and Space Science, Graduate School of Science, Osaka University, Toyonaka, Osaka 560-0043, Japan}
\author{Naoki Koshimoto}
\affiliation{Department of Astronomy, Graduate School of Science, The University of Tokyo, 7-3-1 Hongo, Bunkyo-ku, Tokyo 113-0033, Japan}
\author{Yutaka Matsubara}
\affiliation{Institute for Space-Earth Environmental Research, Nagoya University, Nagoya 464-8601, Japan}
\author{Sho Matsumoto}
\affiliation{Department of Earth and Space Science, Graduate School of Science, Osaka University, Toyonaka, Osaka 560-0043, Japan}
\author{Shota Miyazaki}
\affiliation{Department of Earth and Space Science, Graduate School of Science, Osaka University, Toyonaka, Osaka 560-0043, Japan}
\author{Yasushi Muraki}
\affiliation{Institute for Space-Earth Environmental Research, Nagoya University, Nagoya 464-8601, Japan}
\author{Arisa Okamura}
\affiliation{Department of Earth and Space Science, Graduate School of Science, Osaka University, Toyonaka, Osaka 560-0043, Japan}
\author{Greg Olmschenk}
\affiliation{Code 667, NASA Goddard Space Flight Center, Greenbelt, MD 20771, USA}
\author{Cl\'ement Ranc}
\affiliation{Sorbonne Universit\'e, CNRS, UMR 7095, Institut d'Astrophysique de Paris, 98 bis bd Arago, 75014 Paris, France}
\author{Nicholas J. Rattenbury}
\affiliation{Department of Physics, University of Auckland, Private Bag 92019, Auckland, New Zealand}
\author{Yuki Satoh}
\affiliation{Department of Earth and Space Science, Graduate School of Science, Osaka University, Toyonaka, Osaka 560-0043, Japan}
\author{Takahiro Sumi}
\affiliation{Department of Earth and Space Science, Graduate School of Science, Osaka University, Toyonaka, Osaka 560-0043, Japan}
\author{Daisuke Suzuki}
\affiliation{Department of Earth and Space Science, Graduate School of Science, Osaka University, Toyonaka, Osaka 560-0043, Japan}
\author{Taiga Toda}
\affiliation{Department of Earth and Space Science, Graduate School of Science, Osaka University, Toyonaka, Osaka 560-0043, Japan}
\author{Paul . J. Tristram}
\affiliation{University of Canterbury Mt.\ John Observatory, P.O. Box 56, Lake Tekapo 8770, New Zealand}
\author{Aikaterini Vandorou}
\affiliation{Code 667, NASA Goddard Space Flight Center, Greenbelt, MD 20771, USA}
\affiliation{Department of Astronomy, University of Maryland, College Park, MD 20742, USA}
\author{Hibiki Yama}
\affiliation{Department of Earth and Space Science, Graduate School of Science, Osaka University, Toyonaka, Osaka 560-0043, Japan}
\collaboration{28}{(the MOA Collaboration)}

%

\begin{abstract}
We present the analysis of three more planets from the KMTNet 2021 microlensing season. \onenineteen Lb is a $\sim 6 M_{\rm Jup}$ planet orbiting an early M-dwarf or a K-dwarf, \oneninetwo Lb is a $\sim 2 M_{\rm Nep}$ planet orbiting an M-dwarf, and \twotwoninefour Lb is a $\sim 1.25 M_{\rm Nep}$ planet orbiting a very--low-mass M dwarf or a brown dwarf. These by-eye planet detections provide an important comparison sample to the sample selected with the AnomalyFinder algorithm, and in particular, \twotwoninefour\, is a case of a planet detected by-eye but not by-algorithm. \twotwoninefour Lb is part of a population of microlensing planets around very-low-mass host stars that spans the full range of planet masses, in contrast to the planet population at $\lesssim 0.1\, $ au, which shows a strong preference for small planets.
\end{abstract}


{\section{Introduction} \label{sec:intro}}

This paper is the third in our series that aims to publish all by-eye planet detections from the 2021 Korea Microlensing Telescope Network \citep[KMTNet;][]{kim16} observing season. Previously, in \citet{ryu22} and \citet{ryu22b}, we published 8 planet detections; 10 other planets from the 2021 season have been published in other work \citep{han22a,han22b,han22c,han22d,han22e,yang22}. Here we present the analysis of three additional planetary events: \onenineteen, \oneninetwo, and \twotwoninefour.

The three planetary events were identified by IGS (the first author of this paper) using the traditional ``by-eye'' selection (described in \citealt{ryu22}). However, because IGS uses a variation on method that was previously described, we document that here.  One key element of IGS's selection process is to use an automatic program to fit single-lens/single-point-source (1L1S) light curves \citep{paczynski86} to all events. The 1L1S curves play a key role in providing a reference for noticing anomalies in the observed light curves. Once anomaly-like features are found, IGS conducts initial modeling to reveal what kind of a lens system produces the features. Then, it is possible to decide the selection based on the initial model parameters of the mass ratio and event timescale. For planetary events, the mass ratio should be $\mathcal{O}(10^{-3})$ or smaller, or the event timescale should be shorter than $\sim 10$ days for a relatively small mass ratio (i.e., $\mathcal{O}(10^{-2})$). 

The automatic 1L1S--fitting step is almost identical to the first step of the AnomalyFinder \citep{zang21}. Ultimately, for rigorous statistical analysis, machine-based selection is required to find a well-defined sample of planets. However, there are certain advantages in by-eye selections. First, the human decision process can be used to identify advanced criteria to improve machine-based selection. In addition, because the anomalies are identified based on the insight and experience of a researcher, by-eye selection provides an important cross-check of the algorithm, and in particular in identifying planets that might be missed by an algorithm. In fact, as we will see, the signal in \twotwoninefour\, does not meet the detection criteria of the AnomalyFinder algorithm, which gives us an opportunity to consider the algorithm's failure modes.  
{\section{Observations} \label{sec:obs}}

These planetary microlensing events from 2021 (\onenineteen, \oneninetwo, \twotwoninefour) were first discovered by the Korea Microlensing Telescope Network \citep[KMTNet:][]{kim16} using the KMT AlertFinder \citep{kim18}\footnote{KMTNet Alert System (\url{https://kmtnet.kasi.re.kr/~ulens/})} on (2021 Mar 25, 2021 Mar 29, 2021 Aug 27). The KMTNet observations are made using three identical $1.6$ m telescopes with wide-field cameras (i.e., $2^{\circ}\times2^{\circ}$ field of view).  The telescope network consists of three sites in well-separated timezones, which are located at the Cerro Tololo Inter-American Observatory in Chile (KMTC), the South African Astronomical Observatory in South Africa (KMTS), and the Siding Spring Observatory in Australia (KMTA).  In Table \ref{table:obs}, we summarize observational information of each event. We note that, for KMT-2021-BLG-0192, the Microlensing Observations in Astrophysics \citep[MOA:][]{bond01,sumi03} independently detected the identical event (i.e., MOA-2021-BLG-080 on 2021 Apr 10). Thus, we incorporate the MOA observations in the analysis. 

\begin{deluxetable*}{lccc}
\tablecaption{Observations of 2021 Planetary Events \label{table:obs}}
\tablewidth{0pt}
\tablehead{
\multicolumn{1}{c}{Event} &
\multicolumn{1}{c}{KMT-2021-BLG-0119} &
\multicolumn{1}{c}{KMT-2021-BLG-0192} &
\multicolumn{1}{c}{KMT-2021-BLG-2294} 
}
\startdata
R.A. (J2000)                   & $  18^{h} 16^{m} 00^{s}.13 $        & $  17^{h} 52^{m} 25^{s}.19 $         & $  18^{h} 00^{m} 14^{s}.98 $       \\
Dec (J2000)                    & $ -29^{\circ} 44{'} 38{''}.00 $     & $ -30^{\circ} 00^{'} 31^{''}.28 $    & $ -28^{\circ} 36^{'} 44^{''}.78 $  \\
$(\ell, b)$                    & $ (2^{\circ}.572, -6^{\circ}.155) $ & $ (-0^{\circ}.158, -1^{\circ}.821) $ & $ (1^{\circ}.908, -2^{\circ}.597)$ \\
KMTNet field                   & BLG33                               & BLG02, BLG42                         & BLG03, BLG43                       \\
$\Gamma$ (${\rm hr}^{-1}$)     & 1.0                                 & 4.0                                  & 4.0                                \\
Extinction ($A_{I}$)           & 0.38                                & 2.06                                 & 1.21                               \\
Alert date (YYYY--MM--DD)      & 2021--03--25                        & 2021--03--29                         & 2021--08--27                       \\
Additional Observations        & \nodata                             & MOA                                  & \nodata                            \\
\enddata
\tablecomments{
The MOA alerted MOA-2021-BLG-080 on 2021--04--10, which is the identical event to KMT-2021-BLG-0192.
}
\end{deluxetable*}

For selected events (i.e., planet candidates), individual KMTNet data were carefully re-reduced using photometry packages that adopted the differential image analysis (DIA) technique called pySIS \citep{albrow09} and pyDIA \citep{albrow17,bramich13}. We analyze the light curves using these tender-loving care (TLC) versions of datasets. The MOA data were reduced by their pipeline adopting the DIA method, which is described in \citet{bond01}.

\section{Light Curve Analysis Methodology} \label{sec:basics}

\subsection{Heuristic Analysis} \label{sec:Heuristic}

A planetary microlensing event usually shows a short-term/localized anomaly in the 1L1S light curve. A 1L1S light curve can be described using three parameters: ($t_0, u_0, t_{\rm E}$). These are the time at the peak of the light curve ($t_0$), the impact parameter ($u_0$) in units of the angular Einstein ring radius ($\theta_{\rm E}$), and the Einstein timescale ($t_{\rm E}$), i.e., the travel time of the source through the angular Einstein ring radius. To explain the anomaly induced by a planet, three additional parameters ($s$, $q$, $\alpha$) are required. These are the projected separation between binary components in units of $\theta_{\rm E}$ ($s$), the mass ratio of binary components defined as $q \equiv M_{\rm secondary}/M_{\rm primary}$, and the angle between source trajectory and binary axis ($\alpha$).

From a localized anomaly, we can predict solution(s) using the unified $s^{\dagger}$ formalism described in \citet{hwang22} and \citet{ryu22}. From time of the anomaly, $t_{\rm anom}$, we obtain the scaled time offset from the peak of the light curve, 
\begin{equation}
\tau_{\rm anom} \equiv \frac{t_{\rm anom} - t_0}{t_{\rm E}} 
\end{equation}
and the source position offset from the host,
\begin{equation}
u_{\rm anom} = \sqrt{\tau_{\rm anom}^{2} + u_{0}^{2}} \quad .
\end{equation}
Then, we can also predict,
\begin{equation}
s_{\pm}^{\dagger} \equiv \frac{\sqrt{u_{\rm anom}^{2} + 4} \pm u_{\rm anom}}{2} ;~~ 
\tan \alpha = {\pm} \frac{u_0}{\tau_{\rm anom}},
\end{equation}
where the $\pm$ subscript of $s_{\pm}^{\dagger}$ indicates either a major or minor image perturbation,
respectively \citep{gouldLoeb92}. 
In general, the major--image perturbations ($s_{+}^{\dagger}$) appear as a ``bump''--shaped anomaly, while the minor--image perturbations ($s_{-}^{\dagger}$) appear as a ``dip''--shaped anomaly. For minor--image perturbations, we can additionally predict the mass ratio (to a factor $\sim 2$ level) from the duration of the ``dip" anomaly, $\Delta t_{\rm dip}$ :
\begin{equation}
q = \left( \frac{\Delta t_{\rm dip}}{4 t_{\rm E}} \right)^{2} \frac{s \sin^{2} \alpha}{u_{\rm anom}} 
  = \left( \frac{\Delta t_{\rm dip}}{4 t_{\rm E}} \right)^{2} \frac{s}{|u_{0}|} |\sin^{3} \alpha|.
\end{equation}

The predictions ($s_{\pm}^{\dagger}$) can be compared to the empirical result. In the case of only one solution, $s$ should correspond to one of the values of $s_{\pm}^{\dagger}$. If there are two solutions ($s_{+}$, $s_{-}$), we expect them to be related by 
\begin{equation}
s^{\dagger} = \sqrt{s_{+}s_{-}} \quad .
\end{equation}
In that case, it is the value of $s^{\dagger}$ that should correspond to one of the values of $s_{\pm}^{\dagger}$. The theoretical origins of such degeneracies are discussed in \citet{gaudi97}, \citet{griest98}, and \citet{zhang22}.

\subsection{Basic Modeling} \label{sec:basic_model}

We start the modeling procedure from a static 2L1S model (we treat the static case, i.e., motions of lenses and source are not considered, as a ``standard (STD)'' model. Also, the ``$n$L$m$S'' indicates there are $n$ lenses and $m$ sources), including the finite-source effect, to find the best-fit model describing the observed light curve.  Thus, the STD model requires seven parameters ($t_{0}$, $u_{0}$, $t_{\rm E}$, $s$, $q$, $\alpha$, and $\rho_{\ast}$), where $\rho_{\ast}$ is the angular source radius ($\theta_{\ast}$) scaled by the Einstein radius, i.e., $\rho_{\ast} \equiv \theta_{\ast}/\theta_{\rm E}$. The procedure consists of two basic steps, which may be repeated several times, if necessary.

First, we conduct a grid search to find all possible solutions (i.e., local minima). For the search, we explore $s-q$ parameter space on a grid that spans the values $\log_{10}(s) \in [-1.0, 1.0]$ and $\log_{10}(q) \in [-5.5, 1.0]$. For the remaining parameters ($t_0$, $u_{0}$, $t_{\rm E}$, $\alpha$, and $\rho_{\ast}$), we find optimal solutions using the Markov Chain Monte Carlo (MCMC) algorithm \citep{doran04} to minimize $\chi^2$.  We start the modeling from the 1L1S parameters for $t_0$, $u_0$, and $t_E$, plus $21$ initial values within a range of $\alpha = [0.0,\, 2\pi]$ (radians). We compute the magnification of the 2L1S model using the inverse ray-shooting technique with the ``map-making'' method \citep{dong06, dong09}. Once we find local minima, we explore restricted regions that contain the (possible) local minima, if necessary. 

Second, we refine the possible solutions by setting all parameters to vary freely within (physically) possible ranges. Thus, we obtain fine-tuned model parameters with errors based on the distributions of MCMC chains. During the process of refining the solutions, we rescale the errors of the datasets to make each data point contribute $\chi^{2} \sim 1.0$. We follow the procedure described in \citet{yee12}; i.e., $e_{\rm new} = k\sqrt{e_{\rm old}^2 + e_{\rm min}^2}$, where $e_{\rm new}$ is the rescaled error, $k$ is the rescaling factor, $e_{\rm old}$ is the original error, and $e_{\rm min}$ is the systematics term.

\subsection{Higher-order Effects} \label{sec:high_order}

The STD models assume a static lens system with rectilinear motion relative to the source. However, we should also check for effects from  the observer's motion (i.e., Earth's orbit) or the orbital motion of the binary lens system.

First, we check signals of the annual microlensing parallax \citep[APRX:][]{gould92}, which is caused by the acceleration of Earth. In general, we check the APRX if $t_{\rm E} > 15$ days. For the APRX effect, we introduce two additional parameters: $\pieN$ and $\pieE$, which are north and east components of the microlensing parallax vector ($\pivec_{\rm E}$) projected onto the sky, respectively.  Even if there is not a significant improvement in $\chi^2$, $\pi_{\rm E}$ is often well-constrained along one axis, which is roughly aligned with the $\pieE$ direction. If the APRX model significantly improves $\chi^2$, we investigate the origin of the improvement to check whether or not the APRX measurement is reasonable and not caused by systematics.
  
Second, we also check the lens-orbital (OBT) effect. In reality, the signal caused by the OBT effect is rarely detected. The OBT signal is most often detected in cases with well covered and well-separated (in time) caustic crossings.  Thus, for planetary events (that usually have relatively short anomalies), the OBT effect is hard to detect from the light curve. On the other hand, the OBT effect can affect the APRX measurement because both effects can bend the source trajectory.  Hence, we test the OBT effect to see if it affects the APRX signal. For the OBT effect, we introduce two additional parameters: $ds/dt$ and $d\alpha/dt$, where $ds/dt$ is the rate of change of the binary separation (i.e., $s$) and $d\alpha/dt$ is the rate of change of the $\alpha$ parameter. We also constrain the unphysical solutions using the absolute ratio of transverse kinetic to potential energy \citep{an02, dong09}. That is, by requiring $\beta < 0.8$, where
\begin{equation}
\beta \equiv \left|\frac{\rm KE}{\rm PE}\right|_{\perp} = 
\end{equation}
\begin{equation*}
\left(\frac{\kappa M_{\odot} {\rm yr^{2}}}{8\pi^{2}} \right) \frac{\pi_{\rm E}}{\theta_{\rm E}} 
\left( \frac{s}{\pi_{\rm E} + \pi_{\rm S}/\theta_{\rm E}} \right)^{3} 
\left[ \left( \frac{1}{s} \frac{ds}{dt} \right)^{2} + \left( \frac{d\alpha}{dt} \right)^{2} \right], 
\label{eqn:kepe}
\end{equation*}
where $\kappa$ is a constant deinfed as $\kappa \equiv 4G/c^{2}{\rm au} = 8.144\, {\rm mas}/M_{\odot}$ and $\pi_{\rm S}$ is the source parallax deinfed as $\pi_{\rm S} \equiv {\rm au}/D_{\rm S}$ where $D_{\rm S}$ is the distance to the source.

\subsection{Degenerate Solutions} \label{sec:degeneracy}

We also explicitly check for several types of known degeneracies to be sure we have found all of the relevant 2L1S models and competing solutions.

In addition to the $s^{\dagger}$ (or offset) degeneracy, 2L1S models may be subject to a degeneracy in $\rho_{\ast}$, which may affect the value of $q$ \citep{ryu22,yang22}. Typically, the degeneracy between the two solutions arises because the observed duration of a ``bump" anomaly may be controlled either by the width of the caustic (so $\rho_{\ast}$ is small in comparison) or the size of the source (so $\rho_{\ast}$ is $\gtrsim$ the width of the caustic).  Hence, in some cases, high-cadence observations can distinguish between the two cases, e.g., by demonstrating whether or not the caustic entrance is resolved from the exit.

We also check the 2L1S/1L2S degeneracy \citep{gaudi98}, which \citet{shin19} demonstrated can exist in wider range of cases than those presented in \citet{gaudi98}. This is especially true for light curves that are sparsely covered. For the 1L2S models, we adopt the parameterization described in \citet{shin19} (A--type; see their Appendix), which uses the ratio of the second source flux to the first, $q_{\rm flux}$, and separate values of $t_{0, i}$, $u_{0,i}$, and optionally $\rho_{\ast, i}$ for each source as necessary. Then, we compare the 1L2S model with the best-fit 2L1S solution to see if the 2L1S/1L2S degeneracy can be resolved.

Finally, if we detect the APRX effect, then we check the degenerate APRX solutions, which can be caused by several types such as the ecliptic degeneracy \citep{jiang04, poindexter05}, the ${\pm}u_{0}$ degeneracy \citep{smith03},  and the jerk-parallax Degeneracy \citep{gould04}.\footnote{The APRX degeneracies are well described/organized in \citet{skowron11}.} In practice, the most effective way to find degenerate APRX solutions is to undertake trial searches using different seeds by switching the signs of the parameters: $(u_{0},\, \alpha,\, \pieN) \rightarrow -(u_{0},\, \alpha,\, \pieN)$.

\section{Analysis Results} \label{sec:results}

\subsection{\onenineteen} \label{sec:0119}

\subsubsection{Light Curve}

\begin{figure}[htb!]
\epsscale{1.10}
\plotone{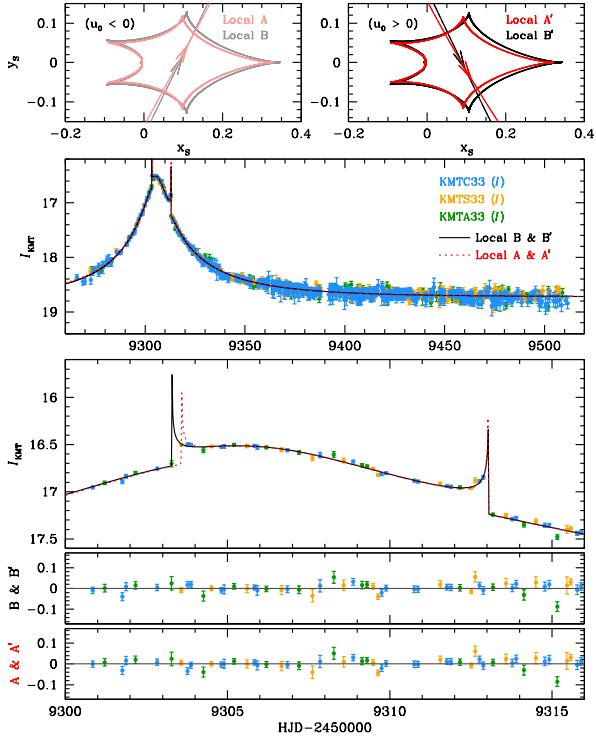}
\caption{Light curve of \onenineteen\ with APRX model curves, geometries, and residuals.
\label{fig:lc_0119}}
\end{figure}

In Figure \ref{fig:lc_0119}, we present the observed light curve of KMT-2021-BLG-0119 (hereafter, KB210119) with the best-fit models (i.e., APRX models) and their caustic geometries. The light curve shows two bump-shaped anomalies.  The anomalies are likely induced by crossings of a central/resonant caustic, which is a potential channel for discovering microlensing planets \citep{han21}.
For the heuristic analysis we have
$(t_{\rm anom} = 9308.3,
t_0 = 9305.97,
u_0 = 0.067,
t_{\rm E} = 62\,$ d)
, which yields
$(u_{\rm anom} = 0.077, 
s^{\dagger}_+ = 1.039,
\alpha = 60\,$ deg).
These values are well-matched to the fitted values derived below.

\subsubsection{STD models and the $\rho_{\ast}$--degeneracy}

For \onenineteen , we find two degenerate families of models. The `A' family of models was found in the standard grid search and the KMTS point at HJD$^{\prime} = 9303.47$ falls on the caustic entrance. In the `B' family of models, which was discovered while checking for $\rho_{\ast} = 0$ solutions, the caustic entrance occurs before this KMTS point. These two families of models have slight differences in the values of microlensing parameters, including $s$ and $q$ (see Table \ref{table:model_0119}).

\begin{deluxetable*}{lrr|rrrr}
\tablecaption{Model Parameters of KMT-2021-BLG-0119 \label{table:model_0119}}
\tablewidth{0pt}
\tablehead{
\multicolumn{1}{c}{Model} &
\multicolumn{2}{c|}{STD} &
\multicolumn{4}{c}{APRX} \\
\multicolumn{1}{c}{Parameter} &
\multicolumn{1}{c}{large$-\rho_{\ast}$} &
\multicolumn{1}{c|}{small$-\rho_{\ast}$} &
\multicolumn{1}{c}{Local A} &
\multicolumn{1}{c}{Local A$^{\prime}$} & 
\multicolumn{1}{c}{Local B} &
\multicolumn{1}{c}{Local B$^{\prime}$}
}
\startdata
$\chi^{2}_{\rm ground}$  & $  1084.207 $ & $  1082.637 $ & $  1060.093 $ & $  1059.787 $ & $  1054.530 $ & $  1054.422 $ \\  
$t_0$ [${\rm HJD'}$]     & $  9305.971 $ & $  9305.956 $ & $  9305.868 $ & $  9305.895 $ & $  9305.769 $ & $  9305.779 $ \\ 
                         & $ \pm 0.032 $ & $ \pm 0.043 $ & $ \pm 0.042 $ & $ \pm 0.043 $ & $ \pm 0.050 $ & $ \pm 0.050 $ \\ 
$u_0$                    & $     0.066 $ & $     0.071 $ & $    -0.077 $ & $     0.076 $ & $    -0.081 $ & $     0.080 $ \\ 
                         & $ \pm 0.002 $ & $ \pm 0.002 $ & $ \pm 0.002 $ & $ \pm 0.003 $ & $ \pm 0.003 $ & $ \pm 0.003 $ \\ 
$t_{\rm E}$ [days]       & $    64.124 $ & $    60.705 $ & $    56.635 $ & $    58.309 $ & $    53.348 $ & $    54.288 $ \\ 
                         & $ \pm 1.523 $ & $ \pm 1.794 $ & $ \pm 1.646 $ & $ \pm 1.793 $ & $ \pm 1.610 $ & $ \pm 1.670 $ \\ 
$s$                      & $     1.039 $ & $     1.043 $ & $     1.049 $ & $     1.045 $ & $     1.054 $ & $     1.053 $ \\ 
                         & $ \pm 0.003 $ & $ \pm 0.003 $ & $ \pm 0.003 $ & $ \pm 0.003 $ & $ \pm 0.003 $ & $ \pm 0.003 $ \\ 
$q$ ($\times 10^{-4}$)   & $    63.607 $ & $    71.886 $ & $    82.413 $ & $    78.592 $ & $    97.455 $ & $    93.935 $ \\ 
                         & $ \pm 3.239 $ & $ \pm 5.612 $ & $ \pm 5.068 $ & $ \pm 5.293 $ & $ \pm 6.149 $ & $ \pm 6.255 $ \\ 
$\alpha$ [rad]           & $     1.045 $ & $     1.046 $ & $    -1.076 $ & $     1.049 $ & $    -1.120 $ & $     1.122 $ \\ 
                         & $ \pm 0.031 $ & $ \pm 0.039 $ & $ \pm 0.028 $ & $ \pm 0.029 $ & $ \pm 0.030 $ & $ \pm 0.030 $ \\ 
$\rho_{\ast}$            & $     0.003 $ & \nodata       &   \nodata     &   \nodata     & \nodata       & \nodata       \\ 
                         & $ \pm 0.001 $ & \nodata       &   \nodata     &   \nodata     & \nodata       & \nodata       \\ 
$\rho_{\ast,{\rm max}}$  & \nodata       & $ < 0.0025  $ & $ < 0.0014  $ & $ < 0.0014  $ & $ < 0.0018  $ & $ < 0.0018  $ \\ 
$\pieN$                  & \nodata       &  \nodata      & $     0.000 $ & $    -0.244 $ & $    -0.031 $ & $    -0.034 $ \\ 
                         & \nodata       &  \nodata      & $ \pm 0.234 $ & $ \pm 0.232 $ & $ \pm 0.222 $ & $ \pm 0.220 $ \\ 
$\pieE$                  & \nodata       &  \nodata      & $     0.143 $ & $     0.132 $ & $     0.209 $ & $     0.183 $ \\ 
                         & \nodata       &  \nodata      & $ \pm 0.032 $ & $ \pm 0.033 $ & $ \pm 0.036 $ & $ \pm 0.037 $ \\ 
\enddata
\tablecomments{
${\rm HJD' = HJD - 2450000.0}$. The total number of data points (${\rm N_{data}}$) is $1059$. 
}
\end{deluxetable*}

\begin{figure}[htb!]
\epsscale{1.10}
\plotone{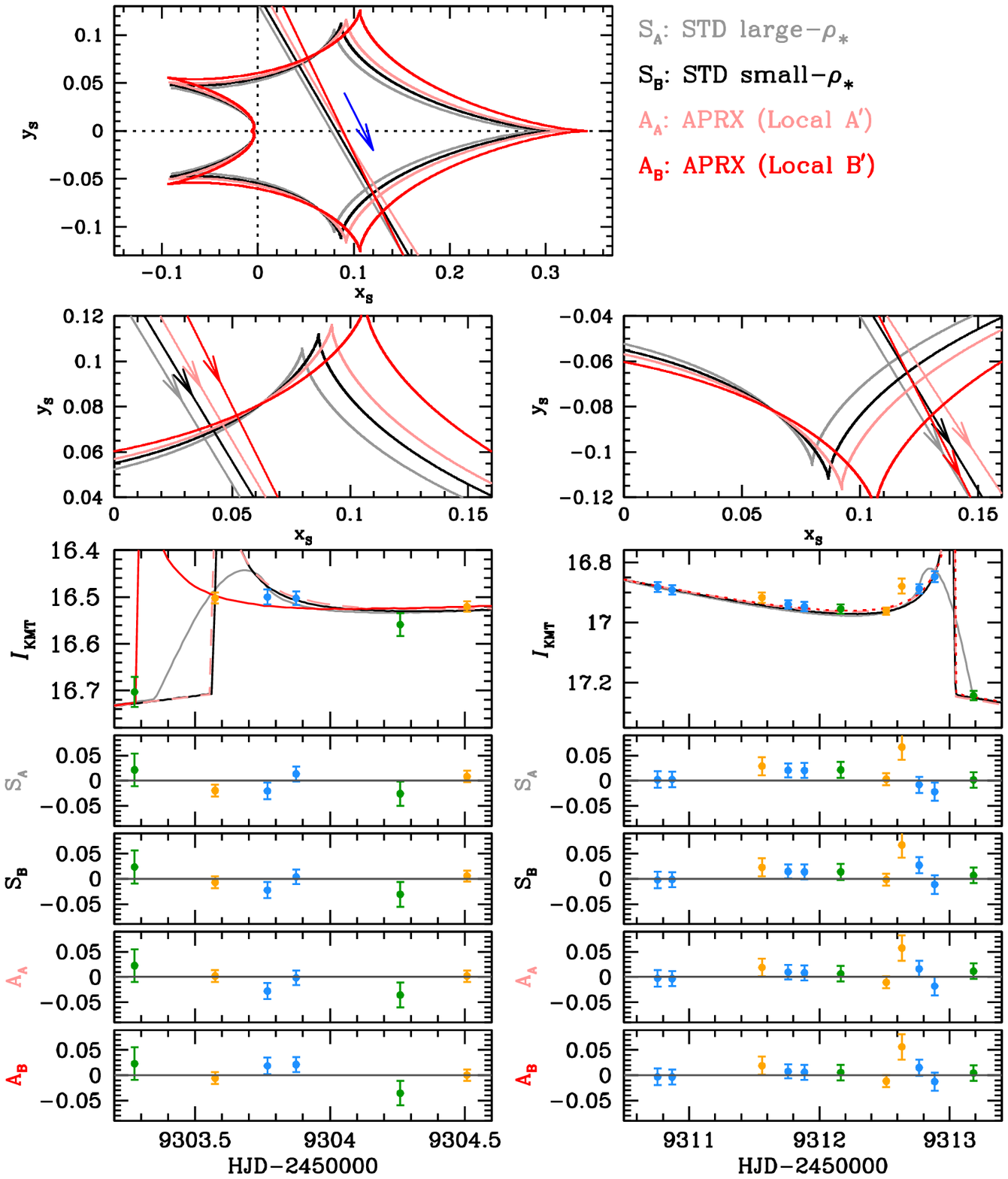}
\caption{The comparison of STD large--$\rho_{\ast}$, STD small--$\rho_{\ast}$, and ARPX models 
of \onenineteen. 
\label{fig:comp_geo_0119}}
\end{figure}

In addition, for the `A' family of models, we find two STD models with very similar values of $s$ and $q$, but different values for $\rho_{\ast}$ (see Table \ref{table:model_0119}).  In Figure \ref{fig:comp_geo_0119}, we present the caustic geometry and the zoom-in on the light curve of each case. The geometries of the two cases are almost identical. However, the observational coverage at the caustic entrance and exit is sub-optimal, so models with both strong finite source effects and no finite source effect fit the data almost equally well. We refer to these as the large--$\rho_{\ast}$ and small--$\rho_{\ast}$ cases, respectively, although the small--$\rho_{\ast}$ case is consistent with $\rho_{\ast} = 0$. We find that the small--$\rho_{\ast}$ case shows better fits at the entrance (i.e., ${\rm HJD^{\prime} = 9303.5 \sim 9304.0}$), while the large--$\rho_{\ast}$ case shows slightly better fits at the exit (i.e., ${\rm HJD^{\prime} = 9312.5 \sim 9313.0}$). We also find that the small--$\rho_{\ast}$ model fits better than the large--$\rho_{\ast}$ fit as the source approaches the caustic exit (i.e., ${\rm HJD^{\prime} = 9311. \sim 9312.5}$). In total, the $\Delta\chi^{2}$ between the large-$\rho_{\ast}$ and small-$\rho_{\ast}$ cases is only $1.57$.

\subsubsection{APRX Models and Solving the $\rho_{\ast}$--degeneracy Problem}

\begin{figure}[htb!]
\epsscale{1.00}
\plotone{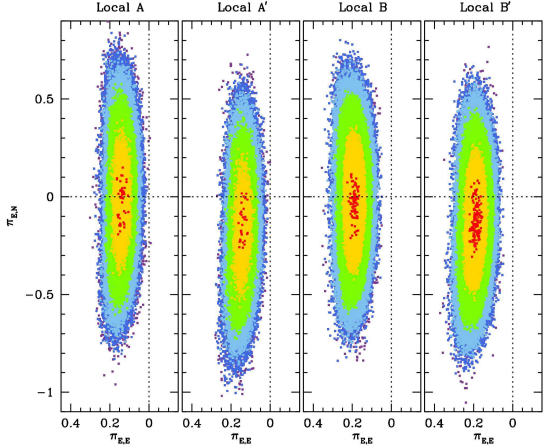}
\caption{The APRX contours of \onenineteen. Each color indicates the $\chi^2$ difference between 
the best-fit and chains. That is, 
$\Delta\chi^{2} \left(\equiv \chi^{2}_{\rm chain} - \chi^{2}_{\rm best-fit}\right) = n^{2}$, 
where $n = 1$ (red), $2$ (yellow), $3$ (green), $4$ (light blue), $5$ (blue), and $6$ (purple).
\label{fig:APRX_0119}}
\end{figure}

The STD models have long timescales ($t_{\rm E} \gtrsim 60$ days), and the two caustic crossings separated by $\sim 9$ days give strong timing constraints on the light curve. Thus, we consider the APRX effect. We find $\chi^2$ improvement $\Delta\chi^{2} = 23 \sim 30$ between the STD (two $\rho_{\ast}$ cases) and APRX ($u_{0} < 0$ and $u_{0} > 0$ cases) models.  In addition, the ARPX contours shown in Figure \ref{fig:APRX_0119} are well converged and inconsistent with zero at $\gtrsim 6\sigma$. 

\begin{figure}[htb!]
\epsscale{1.00}
\plotone{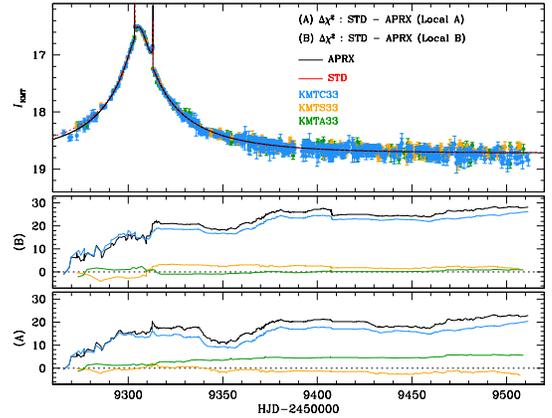}
\caption{Cumulative $\Delta\chi^{2}$ plots between STD and APRX models of \onenineteen.
\label{fig:dchi2_0119}}
\end{figure}

We check the improvements using the cumulative $\Delta\chi^2$ plots shown in Figure \ref{fig:dchi2_0119}. From this investigation, we find that the improvements mostly come from KMTC, which had a higher effective cadence and was taken under better observational conditions than KMTS and KMTA ($3.5$ and $6$ times lower effective cadence). As a result, the contributions of KMTS and KMTA are minor in this case. In addition, the main improvement comes from the left wing, during the beginning of the bulge season when Earth is accelerating rapidly to the East, which can produce the strong $\pieE$ signal as is observed.

\begin{figure}[htb!]
\epsscale{1.10}
\plotone{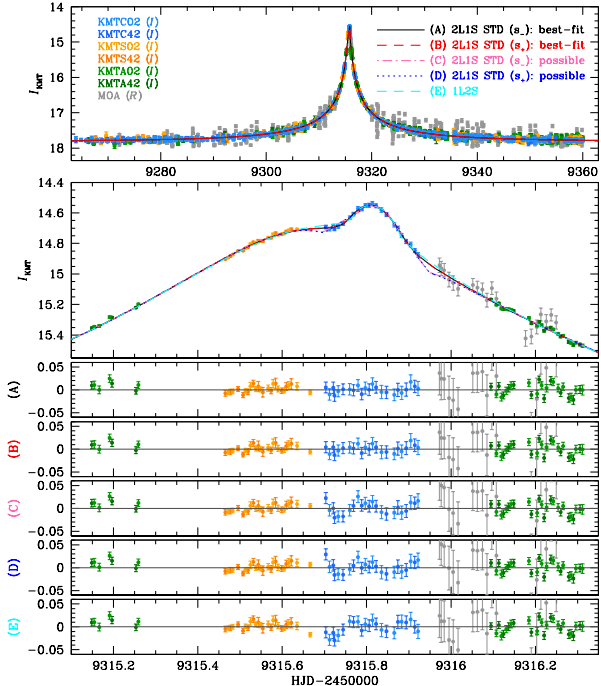}
\caption{The light curve of \oneninetwo\ with STD model curves and residuals. The geometries of 2L1S 
models are presented in Figure \ref{fig:rho_degen_geo_0192}.
\label{fig:lc_0192}}
\end{figure}

Furthermore, we find that, for the `A' family of solutions, APRX models always favor the small--$\rho_{\ast}$ solutions, even when the fits are initialized at the large--$\rho_{\ast}$ STD solutions.  Large--$\rho_{\ast}$ solutions are excluded at the $4\sigma$ level. Indeed, we have a clue about this behavior from STD fits at the caustic--crossings shown in Figure \ref{fig:comp_geo_0119}. The STD models prefer the small--$\rho_{\ast}$ case at the entrance but the large--$\rho_{\ast}$ case at the exit. However, the APRX fits are better than STD fits at both the caustic entrance and exit, including the part approaching the exit. Hence, the $\rho_{\ast}$--degeneracy is resolved when the APRX is included.

\subsubsection{Test of the OBT effect}

We find $\chi^{2}$ improvement of $\Delta\chi^{2} \sim 16$ when we include the OBT parameters in the APRX solutions. However, the OBT parameters show large values, $(ds/dt,\, d\alpha/dt) \sim (0.455,\, -5.973)$, which implies the lens system is unbound or the lens is a very massive object, such as a stellar-mass black hole. If we apply the constraints ${\rm |KE/PE|}_{\perp} < 0.8$ and $M_{\rm L} < 3.0\, M_{\odot}$, we find that most of the $\chi^{2}$ improvement is eliminated. In addition, the OBT parameters are not strongly constrained and are not correlated with the APRX parameters, so we can neglect the OBT in our modeling.

\subsubsection{2L1S/1L2S Degeneracy}

For KB210119, the two planetary anomalies on the light curve are induced by a resonant caustic. Thus, a 1L2S model cannot describe both anomalies. Hence, we do not test the 2L1S/1L2S degeneracy for this event.

\begin{figure}[htb!]
\epsscale{1.10}
\plotone{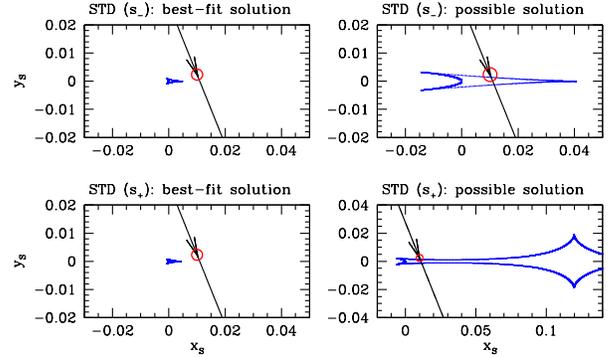}
\caption{Geometries of best--fit and possible solutions ($s_{\pm}$ cases) caused by the $\rho_{\ast}$ 
degeneracy of \oneninetwo.
\label{fig:rho_degen_geo_0192}}
\end{figure}

\subsection{\oneninetwo} \label{sec:0192}

\subsubsection{Heuristic Analysis}

In Figure \ref{fig:lc_0192}, we present the observed light curve of \oneninetwo\ (hereafter, KB210192) with the best-fit STD model. The light curve exhibits a bump anomaly at the peak, which was densely covered by KMTC observations. The localized anomaly has the properties: $\tau_{\rm anom} = 0.00389$ and $u_{\rm anom} = 0.01073$. From the heuristic analysis, we find that $s_{+}^{\dagger} = 1.005$, $s_{-}^{\dagger} = 0.995$ and $\alpha = 1.200$ radians.

\subsubsection{STD Models}

By following procedures described in Section \ref{sec:basic_model}, we conduct STD modeling to find the best-fit models and possible degenerate solutions. We find that there exist two solutions (i.e., $s_{\pm}$ cases) having mass ratios in the planetary event regime, i.e., $q \sim \mathcal{O}(10^{-4})$. The $\Delta\chi^{2}$ between the $s_{-}$ and $s_{+}$ solutions is only $0.510$, so statistically indistinguishable. In Table \ref{table:model_0192}, we present parameters of the best-fit models. The geometric mean of these two solutions is $s^{\dagger} = 1.006$, in good agreement with the $s_{+}^{\dagger}$ prediction from the heuristic analysis. Likewise, the value of $\alpha = 1.184$ is also in good agreement with the heuristic expectation.

\begin{deluxetable*}{lrr|rrrr|lr}
\tablecaption{Model Parameters of KMT-2021-BLG-0192 \label{table:model_0192}}
\tablewidth{0pt}
\tablehead{
\multicolumn{1}{c}{Model} &
\multicolumn{2}{c}{2L1S (STD)} & 
\multicolumn{4}{|c}{2L1S (APRX)} & 
\multicolumn{2}{|c}{1L2S} \\
\multicolumn{1}{c}{Parameter} &
\multicolumn{1}{c}{$s_{-}$} & 
\multicolumn{1}{c}{$s_{+}$} &
\multicolumn{1}{|c}{$s_{-}$ ($u_{0}+$)} & 
\multicolumn{1}{c}{$s_{-}$ ($u_{0}-$)} & 
\multicolumn{1}{c}{$s_{+}$ ($u_{0}+$)} & 
\multicolumn{1}{c}{$s_{+}$ ($u_{0}-$)} &
\multicolumn{1}{|c}{Parameter} &
\multicolumn{1}{c}{} 
}
\startdata
$\chi^{2}$                       & $  4738.844 $ & $  4739.354 $ & $  4702.863 $ & $  4702.032 $ & $  4702.928 $ & $  4702.911 $ & $\chi^{2}$                             & $  4773.792 $ \\  
$t_0$ [${\rm HJD'}$]             & $  9315.697 $ & $  9315.697 $ & $  9315.697 $ & $  9315.697 $ & $  9315.697 $ & $  9315.697 $ & $t_{0,{S_1}}$ [${\rm HJD'}$]           & $  9315.686 $ \\ 
                                 & $ \pm 0.001 $ & $ \pm 0.001 $ & $ \pm 0.001 $ & $ \pm 0.001 $ & $ \pm 0.001 $ & $ \pm 0.001 $ &                                        & $ \pm 0.001 $ \\ 
$u_0$                            & $     0.010 $ & $     0.010 $ & $     0.010 $ & $    -0.010 $ & $     0.010 $ & $    -0.010 $ & $u_{0,{S_1}}$                          & $     0.010 $ \\ 
                                 & $ \pm 0.001 $ & $ \pm 0.001 $ & $ \pm 0.001 $ & $ \pm 0.001 $ & $ \pm 0.001 $ & $ \pm 0.001 $ &                                        & $ \pm 0.001 $ \\ 
$t_{\rm E}$ [days]               & $    32.601 $ & $    32.257 $ & $    31.539 $ & $    31.627 $ & $    31.204 $ & $    31.890 $ & $t_{0,{S_2}}$ [${\rm HJD'}$]           & $  9315.823 $ \\ 
                                 & $ \pm 0.605 $ & $ \pm 0.611 $ & $ \pm 0.716 $ & $ \pm 0.717 $ & $ \pm 0.725 $ & $ \pm 0.696 $ &                                        & $ \pm 0.001 $ \\ 
$s$                              & $     0.776 $ & $     1.303 $ & $     0.774 $ & $     0.761 $ & $     1.321 $ & $     1.312 $ & $u_{0,{S_2}}$ ($\times 10^{-3}$)       & $    -0.058 $ \\ 
                                 & $ \pm 0.016 $ & $ \pm 0.028 $ & $ \pm 0.018 $ & $ \pm 0.017 $ & $ \pm 0.030 $ & $ \pm 0.030 $ &                                        & $ \pm 0.287 $ \\ 
$q$ ($\times 10^{-4}$)           & $     3.327 $ & $     3.333 $ & $     3.541 $ & $     3.733 $ & $     3.707 $ & $     3.544 $ & $t_{\rm E}$ [days]                     & $    32.321 $ \\ 
                                 & $ \pm 0.323 $ & $ \pm 0.326 $ & $ \pm 0.370 $ & $ \pm 0.363 $ & $ \pm 0.362 $ & $ \pm 0.363 $ &                                        & $ \pm 0.623 $ \\ 
$\alpha$ [rad]                   & $     1.184 $ & $     1.184 $ & $     1.183 $ & $    -1.186 $ & $     1.181 $ & $    -1.184 $ & $\rho_{\ast,{S_1},{\rm max}}$          & $   < 0.0087$ \\ 
                                 & $ \pm 0.005 $ & $ \pm 0.004 $ & $ \pm 0.005 $ & $ \pm 0.005 $ & $ \pm 0.005 $ & $ \pm 0.005 $ &                                        &               \\ 
$\rho_{\ast}$ ($\times 10^{-4}$) & $    19.607 $ & $    19.318 $ & $    20.028 $ & $    19.212 $ & $    20.063 $ & $    19.201 $ & $\rho_{\ast,{S_2}}$ ($\times 10^{-4}$) & $    16.277 $ \\ 
                                 & $ \pm 1.634 $ & $ \pm 1.696 $ & $ \pm 1.892 $ & $ \pm 1.824 $ & $ \pm 1.861 $ & $ \pm 1.878 $ &                                        & $ \pm 0.755 $ \\ 
\nodata                          & \nodata       & \nodata       & \nodata       & \nodata       & \nodata       &   \nodata     & $q_{\rm flux}$                         & $     0.019 $ \\
                                 & \nodata       & \nodata       & \nodata       & \nodata       & \nodata       &   \nodata     &                                        & $ \pm 0.001 $ \\
$\pieN$                          & \nodata       & \nodata       & $     1.637 $ & $     2.312 $ & $     2.896 $ & $     2.143 $ & \nodata                                & \nodata       \\
                                 & \nodata       & \nodata       & $ \pm 2.124 $ & $ \pm 2.117 $ & $ \pm 2.143 $ & $ \pm 2.081 $ & \nodata                                & \nodata       \\
$\pieE$                          & \nodata       & \nodata       & $     0.272 $ & $     0.304 $ & $     0.360 $ & $     0.285 $ & \nodata                                & \nodata       \\
                                 & \nodata       & \nodata       & $ \pm 0.140 $ & $ \pm 0.137 $ & $ \pm 0.141 $ & $ \pm 0.135 $ & \nodata                                & \nodata       \\
\enddata
\tablecomments{
${\rm HJD' = HJD - 2450000.0}$. The total number of data points (${\rm N_{data}}$) is $4676$. 
For the 1L2S model, the angular radius of the first source ($\rho_{\ast,{S_1}}$) is not measured. 
The best-fit velue of $\rho_{\ast,{S_1}}$ is $7.690\times10^{-4}$.
}
\end{deluxetable*}

Figure \ref{fig:lc_0192} shows that the best-fit solutions do not have caustic-crossing geometries. However, for this event, the source's proximity to the cusp along the binary axis means that it passes over a relatively sharp magnification ``ridge" that allows a measurement of $\rho_{\ast}$. The extremely dense coverage at the anomaly makes this measurement very secure.

\subsubsection{Resolving the $\rho_{\ast}$ degeneracy}

The caustic geometry of \oneninetwo\, is similar to the cases of KMT-2021-BLG-1391Lb and KMT-2021-BLG-1253Lb \citep{ryu22}, which suggests there may be alternate solutions for \oneninetwo\, caused by the $\rho_{\ast}$ degeneracy. We explicitly search for such solutions and present the caustic geometries of the possible large--$\rho_{\ast}$ solutions compared to those of the best-fit solutions in Figure \ref{fig:rho_degen_geo_0192}. The possible solutions show worse fits with $\Delta\chi^{2} = 23.4$ and $21.7$ for $s_{-}$ and $s_{+}$ the cases, respectively. The caustic-crossing feature cannot describe the observations at the anomaly very well. Thus, because of the extremely dense coverage, we can resolve the $\rho_{\ast}$ degeneracy for this event.

\subsubsection{Resolving the 2L1S/1L2S degeneracy}

Localized bump-shaped anomalies, like that seen in \oneninetwo, may also be explained by a 1L2S interpretation. We find a plausible 1L2S model shown in Table \ref{table:model_0192}. Both the flux ratio of binary sources, $q_{\rm flux} \equiv {\rm flux}_{S_{2}} / {\rm flux}_{S_{1}}$, and $\rho_{\ast, S_{2}}$ are well-measured, but there is only an upper limit on $\rho_{\ast, S_{1}}$, which may be either larger or smaller than $\rho_{\ast, S_{2}}$. Hence, it is not possible to rule out this solution based on these physical considerations. On the other hand, for this solution $\theta_{\ast,2} \sim 0.3\, \mu$as, and $t_{\ast,2} = 0.05$ d, so $\mu = \theta_{\ast}/t_{\ast} = 1.6\, \mathrm{mas\, yr}^{-1}$, which is somewhat unlikely, though not impossible. In addition, the 1L2S solution fits worse than the 2L1S solutions by $\Delta\chi^{2} \sim 35$ (more relative to APRX, see below), so it can be ruled out on that basis.

\subsubsection{Tests of APRX and OBT effects}

Because the timescale of this event is about $1$ month (i.e., $t_{\rm E} \sim 32$ days), it is worth testing the detection of the APRX signal. In our initial fits, we found an extreme value of the parallax with $|\pieN| > 2$. However, our investigation of the cumulative $\Delta\chi^{2}$ plots showed that the $\chi^{2}$ improvement mostly came from the baseline data toward the end of the microlensing season. Thus, we exclude data with ${\rm HJD^{\prime}} > 9360.0$ from the modeling for this event.

\begin{figure}[htb!]
\epsscale{1.00}
\plotone{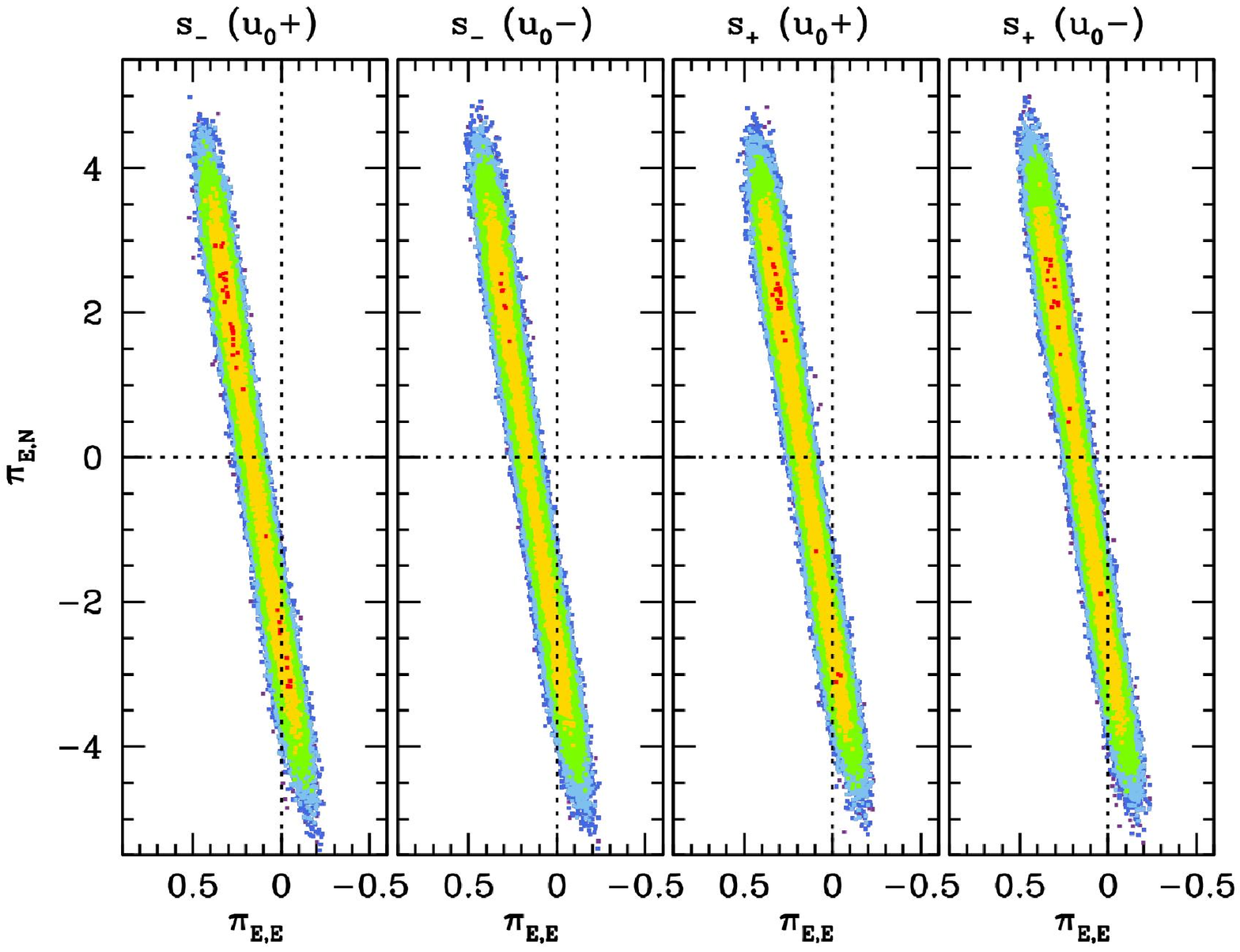}
\caption{The $(\pieE, \pieN)$ distributions of APRX models of \oneninetwo. The color scheme is identical 
to Figure \ref{fig:APRX_0119}.
\label{fig:APRX_0192}}
\end{figure}

Ultimately, we find that the parallax improves the fit by $\Delta\chi^2 \sim 37$. We present the APRX distributions in Figure \ref{fig:APRX_0192}. While the magnitude of the parallax is not well-constrained, the vector is well-constrained along one axis (as expected).

In addition, we check for the OBT effect. The APRX+OBT models strongly prefer unphysical values for the OBT parameters (implying unbound orbits). However, including the OBT parameters does not affect the parallax constraints. Therefore, we suppress OBT effects in our modeling.

\subsection{\twotwoninefour} \label{sec:2294}

\subsubsection{Heuristic Analysis}

\begin{figure}[htb!]
\epsscale{1.00}
\plotone{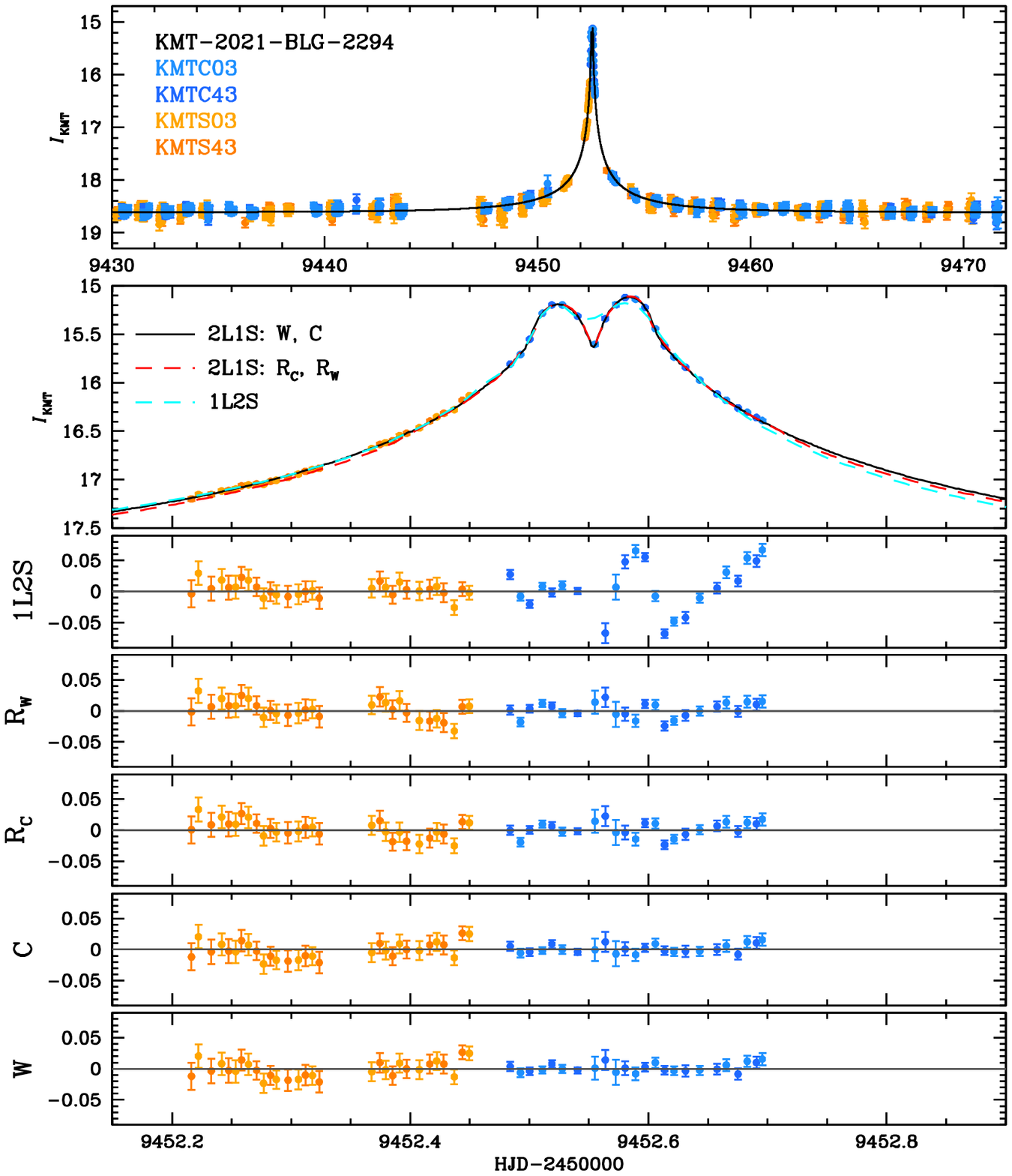}
\caption{Light curve of \twotwoninefour\ with model curves and their residuals. The geometries of 2L1S 
models are presented in Figure \ref{fig:geo_2294}.
\label{fig:lc_2294}}
\end{figure}

The light curve of \twotwoninefour\ shows a clear anomaly at the peak of the light curve (see Figure \ref{fig:lc_2294}). Because the anomaly occurs at the peak of the event $\tau_{\rm anom} \sim 0$ and $u_{\rm anom} \sim u_0 = 0.006$. Hence, the heuristic analysis suggests $s^{\dagger}_{-} = 0.997$ and $s^{\dagger}_+ = 1.003$ and $\alpha = \pm\pi/2$ radians. Furthermore, because this is a ``dip" anomaly, we can predict the mass ratio from $\Delta t_{\rm dip} = 0.06\, $ days and $t_{\rm E} = 7.1\,$ days; i.e., $q = 7.4 \times 10^{-4}$.

\subsubsection{STD Models}

The KMTA images have extremely low S/N for the source and did not cover the anomaly or other magnified parts of the light curve (there are no data from ${\rm HJD^{\prime}} \sim 9451$ to $\sim 9454$) Thus, we do not include KMTA data in the modeling. 

\begin{figure}[htb!]
\epsscale{1.00}
\plotone{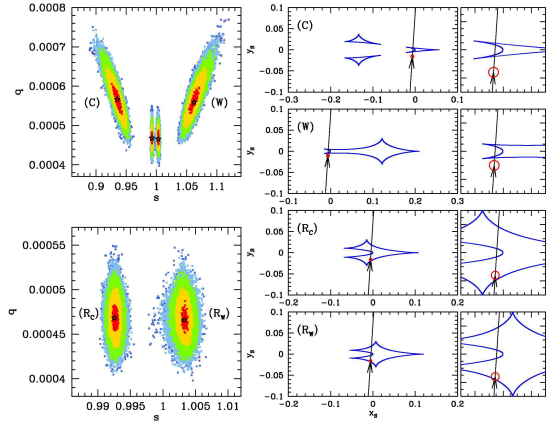}
\caption{The 2L1S model geometries of the best--fit and possible solutions (right-side panels) of 
\twotwoninefour. The left-side panels show the locations of the solutions in the $(s,q)$ parameter 
space. The color scheme of the space is identical to Figure \ref{fig:APRX_0119}.
\label{fig:geo_2294}}
\end{figure}

From the detailed modeling, we find that there exist four degenerate solutions. Figure \ref{fig:geo_2294} shows four solutions in $s-q$ parameter space and also presents the caustic geometry of each solution. Their best-fit model parameters are given in Table \ref{table:model_2294}.

\begin{deluxetable*}{lrrrr}
\tablecaption{Model Parameters of KMT-2021-BLG-2294 \label{table:model_2294}}
\tablewidth{0pt}
\tablehead{
\multicolumn{1}{c}{Parameter} &
\multicolumn{1}{c}{Close (C)} &
\multicolumn{1}{c}{Resonant (${\rm R_C}$)} &
\multicolumn{1}{c}{Resonant (${\rm R_W}$)} &
\multicolumn{1}{c}{Wide (W)} 
}
\startdata
$\chi^{2}_{\rm ground} / {\rm N_{data}}$  & $ 8219.429 / 8254    $ & $ 8243.348 / 8254    $ & $ 8248.935 / 8254    $ & $ 8218.934 / 8254    $ \\  
$t_0$ [${\rm HJD'}$]                      & $ 9452.558 \pm 0.001 $ & $ 9452.558 \pm 0.001 $ & $ 9452.558 \pm 0.001 $ & $ 9452.558 \pm 0.001 $ \\ 
$u_0$                                     & $    0.006 \pm 0.001 $ & $    0.005 \pm 0.001 $ & $    0.005 \pm 0.001 $ & $    0.006 \pm 0.001 $ \\ 
$t_{\rm E}$ [days]                        & $    7.074 \pm 0.253 $ & $    8.067 \pm 0.290 $ & $    8.038 \pm 0.297 $ & $    7.144 \pm 0.258 $ \\ 
$s$                                       & $    0.935 \pm 0.009 $ & $    0.993 \pm 0.001 $ & $    1.003 \pm 0.001 $ & $    1.062 \pm 0.010 $ \\ 
$q$ ($\times 10^{-3}$)                    & $    0.567 \pm 0.041 $ & $    0.468 \pm 0.018 $ & $    0.466 \pm 0.018 $ & $    0.559 \pm 0.040 $ \\ 
$\alpha$ [rad]                            & $    4.777 \pm 0.006 $ & $    4.776 \pm 0.005 $ & $    4.777 \pm 0.005 $ & $    4.777 \pm 0.005 $ \\ 
$\rho_{\ast}$                             & $    0.003 \pm 0.001 $ & $    0.003 \pm 0.001 $ & $    0.003 \pm 0.001 $ & $    0.003 \pm 0.001 $ \\ 
\enddata
\tablecomments{
${\rm HJD' = HJD - 2450000.0}$. 
}
\end{deluxetable*}

The degeneracies arise from a combination of the $s_{\pm}$ degeneracy and an unexpected resonant caustic degeneracy. We refer to the preferred set of solutions as ``C'' (close) and ``W'' (wide). The ``W" solution has a resonant caustic, but the ``C" solution does not. For these solutions, $s^{\dagger} = 0.996$ and $\alpha = 4.777$ radians, in good agreement with the heuristic analysis. The $\Delta\chi^{2}$ between the best-fit (i.e., ``W'' case) and the ``C" case is only $0.5$.
The close and wide cases produce almost identical 
light curves and so are completely degenerate. 

The second pair of solutions both have resonant caustics, so we refer to them as ``${\rm R_C}$'' (Resonant, $s<1$) and ``${\rm R_W}$'' (Resonant, $s>1$). These also obey the expectations from the heuristic analysis with $s^{\dagger} = 0.998$ and $\alpha = 4.777$. One remarkable aspect about these solutions is that $\rho$ is very similar to the ``C" and ``W" solutions. Examining the source trajectory and caustic structure in Figure \ref{fig:geo_2294} suggests that there should be four distinct caustic crossings even though only two bumps are seen in the light curve. In fact, due to the source location at the outer edges of the caustics, those crossings (which would occur at ${\rm HJD}^{\prime} = 9452.39$ and $9452.71$) are so weak as to produce almost no change in magnification relative to a point lens. Nevertheless, these slight differences lead to these solutions being disfavored relative to the ``W" case by $24.4$ (``${\rm R_C}$'') and $30.0$ (``${\rm R_W}$'').

\subsubsection{Tests of ARPX and OBT effects}

Because of the short timescale of this event (i.e., $7\sim8$ days), we do not attempt to place limits on APRX or OBT effects.

\subsubsection{2L1S/1L2S Degeneracy}

The feature at the peak might seem to be explainable by a 1L2S interpretation. However, we find that the 1L2S models cannot describe the peak of the light curve, and especially not the KMTC03 point at HJD$' = 9452.55$. In total, the 1L2S model is disfavored by $\Delta\chi^{2} \sim 770$ relative to the planetary models.

\section{Source Color and Angular Source Radius} \label{sec:cmd}

When $\rho_{\ast}$ is measured, it can be used to determine the angular Einstein ring radius ($\theta_{\rm E} = \theta_{\ast} / \rho_{\ast}$, where $\theta_{\ast}$ is the angular source radius). While $\rho_{\ast}$ was measured for \oneninetwo\ and \twotwoninefour, for \onenineteen, we can only measure the $\rho_{\ast}$ distribution, which can be used to set limits on $\theta_{\rm E}$ in the Bayesian analysis (Section \ref{sec:lens_properties}).

\begin{figure*}[htb!]
\epsscale{1.00}
\plotone{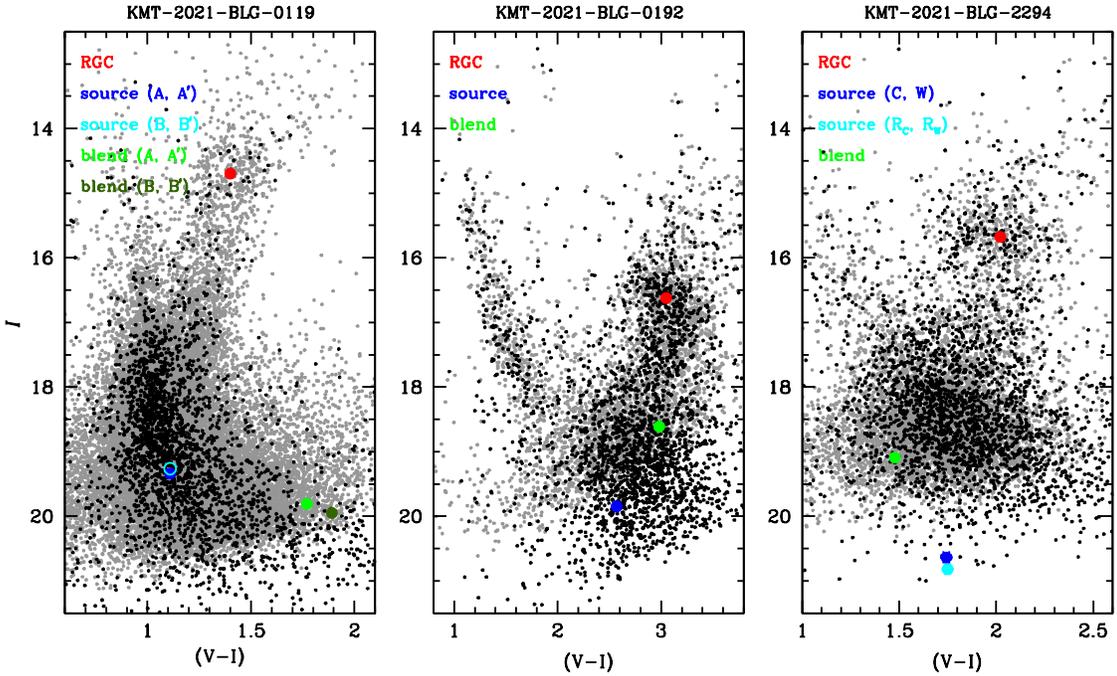}
\caption{The color--magnitude diagrams of three events. 
The color and magnitudes of the KMTNet CMD (black dot) are aligned to the OGLE--III (gray dot) 
instrumental scales. The colored circles indicate the positions of RGC (red), source (blue/cyan), 
and blend(green/dark green) shown in Table \ref{table:cmd}.
\label{fig:CMDs}}
\end{figure*}

We measure $\theta_{\ast}$ for all events using the conventional method described in \citet{yoo04}. In Figure \ref{fig:CMDs}, we present the $V$/$I$ CMD of each event with the of the centroid of the red giant clump (RGC), source, and blend. In Table \ref{table:cmd}, we present the results of the CMD analyses. The intrinsic color of the RGC is adopted from \citet{bensby11}. The de-reddened magnitude of the RGC is adopted from \citet{nataf13} according to the galactic longitude of each event. Under the assumption that the source and RGC experienced the same stellar extinction, we can obtain the de-reddened color and magnitude of the source. Based on the source color, we determine $\theta_{\ast}$ using the surface brightness--color relation of \citet{kervella04}\footnote{Because \citet{kervella04} provide the relation based on the ($V-K$) color, we convert the source color from ($V-I$) to ($V-K$) using the color conversion of \citet{BB88}}. 

\begin{deluxetable*}{lccc}
\tablecaption{CMD analysis of Three Events \label{table:cmd}}
\tablewidth{0pt}
\tablehead{
\multicolumn{1}{c}{Event} &
\multicolumn{1}{c}{KMT-2021-BLG-0119} &
\multicolumn{1}{c}{KMT-2021-BLG-0192} &
\multicolumn{1}{c}{KMT-2021-BLG-2294} 
}
\startdata
$(V-I, I)_{\rm cl}$             & $(1.40, 14.70)$                   & $(3.05, 16.62)$                         & $(2.02, 15.68)$                      \\
$(V-I, I)_{0, \rm cl}$          & $(1.06, 14.36)$                   & $(1.06, 14.45)$                         & $(1.06, 14.38)$                      \\
\hline
Solution                        & Local A \& A$^{\prime}$           & STD ($s_{-}$), APRX ($s_{\pm}, u{0}+$)  & C, W                                 \\
$(V-I, I)_{\rm S}$              & $(1.11\pm0.02, 19.34\pm0.01)$     & $(2.57\pm0.01, 19.85\pm0.01)$           & $(1.84 \pm 0.01, 20.72 \pm 0.01)$    \\
$(V-I, I)_{0, \rm S}$           & $(0.77\pm0.05, 19.00\pm0.01)$     & $(0.58\pm0.05, 17.68\pm0.01)$           & $(0.88 \pm 0.05, 19.42 \pm 0.01)$    \\
$(V-I, I)_{\rm B}$              & $(1.77\pm0.09, 19.81\pm0.02)$     & $(2.98\pm0.02, 18.61\pm0.01)$           & $(1.58 \pm 0.02, 19.18 \pm 0.01)$    \\
$\theta_{\ast}$ ($\mu{\rm as}$) & $ 0.53\pm0.03 $                   & $ 0.80\pm0.04 $                         & $ 0.50 \pm 0.03 $                    \\
$\theta_{\rm E}$ (mas)          & $ > 0.38$                         & $ 0.40\pm0.05 $                         & $ 0.15 \pm 0.02 $                    \\
\hline                                                                                                  
Solution                        & Local B \& B$^{\prime}$           & STD ($s_{+}$), APRX ($s_{\pm}, u{0}-$)  & ${\rm R}_{\rm C}$, ${\rm R}_{\rm W}$ \\
$(V-I, I)_{\rm S}$              & $(1.11\pm0.02, 19.26\pm0.01)$     & $(2.57\pm0.01, 19.84\pm0.01)$           & $(1.85 \pm 0.01, 20.90 \pm 0.01)$    \\
$(V-I, I)_{0, \rm S}$           & $(0.77\pm0.05, 18.92\pm0.01)$     & $(0.58\pm0.05, 17.68\pm0.01)$           & $(0.89 \pm 0.05, 19.60 \pm 0.01)$    \\
$(V-I, I)_{\rm B}$              & $(1.89\pm0.12, 19.95\pm0.02)$     & $(2.98\pm0.02, 18.61\pm0.01)$           & $(1.59 \pm 0.02, 19.14 \pm 0.01)$    \\
$\theta_{\ast}$ ($\mu{\rm as}$) & $ 0.55\pm0.03 $                   & $ 0.80\pm0.04 $                         & $ 0.47 \pm 0.03 $                    \\
$\theta_{\rm E}$ (mas)          & $> 0.31$                          & $ 0.42\pm0.05 $                         & $ 0.17 \pm 0.02 $                    \\
\enddata
\end{deluxetable*}

We note that, for \onenineteen, the red giant stars in the KMTNet CMD are too sparse to precisely determine the RGC. Thus, we use the OGLE-III CMD \citep{szymanski11} to determine the RGC. The instrumental color and magnitude of KMTNet are aligned to the OGLE instrumental scales using the cross--matching of field stars. For the other events, the RGC can be determined from the KMTNet CMDs. However, for consistency, we present the results of the CMD analyses scaled to OGLE-III.

\section{Characteristics of Planets} \label{sec:lens_properties}

\subsection{Bayesian Analysis} \label{sec:basic_Bayes}

For the Bayesian analysis, we adopt the formalism described in \citet{shin21}, except that we adopt initial and present-day mass functions from \citet{chabrier03}. In brief, we adopt the other Galactic priors from several studies:
\begin{enumerate}
    \item {the matter density profile of the disk from 
\citet{robin03} and \citet{bennett14},}
    \item{the matter density profile of the bulge from \citet{hangould95} and \citet{dwek95},}
    \item{the mean velocity and the velocity dispersion of bulge stars from GAIA proper motion information \citep{gaia18}, and}
    \item{the mean velocity and the velocity dispersion of disk stars from the modified model of \citet{hangould95}, which is described in \citet{han20}.}
\end{enumerate}

We then generate artificial microlensing events (total $4\times10^{7}$ events for each case) and apply available constraints from the microlensing light curve. For all cases, $t_{\rm E}$ are well measured, so we use a simple Gaussian weight. Depending on the particular event, we may also have priors from $\theta_{\rm E}$ or $\pi_{\rm E}$. For \oneninetwo\ and \twotwoninefour, we measure $\rho_{\ast}$, so we apply a Gaussian weight based on $\theta_{\rm E}$ (see Section \ref{sec:cmd}). In addition, for \oneninetwo, apply the 2D $\pi_{\rm E}$ constraint following the formalism described in \citet{ryu19}.

For \onenineteen, we also use the 2D APRX distributions as a constraint for $\pi_{\rm E}$. Then, because $\rho_{\ast}$ was not clearly measured, for each solution, we construct a weight function ($W(\rho_{\ast})$) by fitting of the distribution of $\Delta\chi^2$ as a function of $\rho_{\ast}$. For the Local A and A$^{\prime}$ cases, we use a piece-wise
function 
\begin{equation}
W(\rho_{\ast}) = 
\begin{cases}
        {\rm constant}                & {( \rm if~ \rho_{\ast} < \rho_{\ast, {\rm break}} )} \\
a\, e^{c(x+b)^{2}} + 0.6 & {( \rm if~ \rho_{\ast, {\rm break}} \le \rho_{\ast} \le \rho_{\ast, {\rm limit}} )} \\
        0.0                           & {( \rm if~ \rho_{\ast} > \rho_{\ast, {\rm limit}} )} 
\end{cases}
,
\end{equation}
where $x \equiv \log_{10}(\rho_{\ast})$ and ($a$, $b$, $c$) are coefficient set for fitting, and $0.6$ is the normalization factor for making unity weight at the best-fit value. For the Local A and A$^{\prime}$ cases, $\rho_{\ast}$ cannot be zero because there is a point during the caustic entrance. However, it is increasingly difficult to probe models with $\rho_{\ast} < \rho_{\rm \ast, break}$ through an MCMC (which tends toward the preferred B and B$^{\prime}$ cases). At the same time, these values are increasingly unlikely because they imply ever larger values of $\theta_{\rm E}$ ($\theta_{\rm E} (\rho_{\rm \ast, break} = 6.6\times10^{-5}) \sim 80\, $ mas), so our assumption of a constant weight below $\rho_{\rm \ast, break}$ has little effect on the Bayesian estimates. For the Local B and B$^{\prime}$ cases, we use 
\begin{equation}
W(\rho_{\ast}) =
\begin{cases}
a\, e^{\frac{x-b}{c}} + 1.0 & {( \rm if~ \rho_{\ast} \le \rho_{\ast, {\rm limit}} )} \\
                        0.0              & {( \rm if~ \rho_{\ast} > \rho_{\ast, {\rm limit}} )} 
\end{cases}
,
\end{equation}
where $x \equiv \log_{10}(\rho_{\ast})$, ($a$, $b$, $c$) are coefficients, and $1.0$ is the normalization factor. We present the coefficients for all models, 
$\rho_{\ast, {\rm limit}}$, and $\rho_{\ast, {\rm break}}$ in Table \ref{table:coeff_0119}.

\begin{deluxetable}{lrrrr}
\tablecaption{Coefficients of the $\rho_{\ast}$ weight functions for \onenineteen \label{table:coeff_0119}}
\tablewidth{0pt}
\tablehead{
\multicolumn{1}{c}{Coefficient} &
\multicolumn{1}{c}{Local A} &
\multicolumn{1}{c}{Local A$^{\prime}$} &
\multicolumn{1}{c}{Local B} &
\multicolumn{1}{c}{Local B$^{\prime}$} 
}
\startdata
a                         &  0.385439 &  0.402486 & -59.035155 & -55.573560 \\ 
b                         &  3.811082 &  3.899674 &  -0.497434 &  -0.688004 \\
c                         & -4.760185 & -3.133795 &   0.414940 &   0.393204 \\
$\rho_{\ast,{\rm limit}}$ &  0.002344 &  0.002512 &   0.002344 &   0.002042 \\
$\rho_{\ast,{\rm break}}$ &  0.000066 &  0.000054 & \nodata    & \nodata    \\
\enddata
\end{deluxetable}

\begin{deluxetable*}{lclrrrrrrr}
\tablecaption{Lens Properties of Three Events \label{table:lens}}
\tablewidth{0pt}
\tablehead{
\multicolumn{1}{c}{Event}                 &
\multicolumn{1}{c}{Constraints}           &
\multicolumn{1}{c}{Case}                  &
\multicolumn{1}{c}{$M_{\rm host}$}        &
\multicolumn{1}{c}{$M_{\rm planet}$}      &
\multicolumn{1}{c}{$M_{\rm planet}$}      &
\multicolumn{1}{c}{$D_{\rm L}$}           &
\multicolumn{1}{c}{$a_{\perp}$}           &
\multicolumn{1}{c}{$a_{\rm snow}$}        &
\multicolumn{1}{c}{$\mu_{\rm rel}$}       \\
\multicolumn{1}{c}{}                      &
\multicolumn{1}{c}{}                      &
\multicolumn{1}{c}{}                      &
\multicolumn{1}{c}{($M_{\odot}$)}         &
\multicolumn{1}{c}{($M_{\rm Jup}$)}       &
\multicolumn{1}{c}{($M_{\rm Nep}$)}       &
\multicolumn{1}{c}{(kpc)}                 &
\multicolumn{1}{c}{(au)}                  &
\multicolumn{1}{c}{(au)}                  &
\multicolumn{1}{c}{($\rm mas\, yr^{-1}$)}
}
\startdata
KB210119 & $t_{\rm E} + \rho_{\ast} + \pivec_{\rm E}$    & Local A              & $ 0.69_{-0.30}^{+0.34} $ & $ 5.97_{-2.60}^{+2.94} $ & $ 110.8_{-48.2}^{+54.6} $ & $ 3.51_{-1.13}^{+1.72} $ & $ 3.23_{-0.80}^{+0.76} $ & $ 1.87_{-0.80}^{+0.92} $ & $ 5.86_{-2.55}^{+3.15} $ \\ 
         &                                               & Local A$^{\prime}$   & $ 0.69_{-0.30}^{+0.34} $ & $ 5.67_{-2.53}^{+2.87} $ & $ 105.2_{-46.9}^{+53.2} $ & $ 3.69_{-1.20}^{+1.75} $ & $ 3.24_{-0.87}^{+0.78} $ & $ 1.86_{-0.82}^{+0.93} $ & $ 5.39_{-2.34}^{+3.07} $ \\
         &                                               & Local B              & $ 0.55_{-0.23}^{+0.31} $ & $ 5.58_{-2.47}^{+3.16} $ & $ 103.4_{-45.8}^{+58.5} $ & $ 3.05_{-0.91}^{+1.29} $ & $ 2.87_{-0.67}^{+0.67} $ & $ 1.47_{-0.63}^{+0.83} $ & $ 6.29_{-2.56}^{+3.18} $ \\ 
         &                                               & Local B$^{\prime}$   & $ 0.56_{-0.24}^{+0.32} $ & $ 5.52_{-2.40}^{+3.12} $ & $ 102.4_{-44.5}^{+57.9} $ & $ 3.13_{-0.91}^{+1.30} $ & $ 2.92_{-0.68}^{+0.67} $ & $ 1.51_{-0.64}^{+0.85} $ & $ 6.11_{-2.46}^{+3.09} $ \\ 
\hline
KB210192 & $t_{\rm E} + \theta_{\rm E}$                  & $s_{-}$              & $ 0.55_{-0.28}^{+0.26} $ & $ 0.19_{-0.10}^{+0.09} $ & $ 3.55_{-1.80}^{+1.75} $ & $ 6.66_{-1.41}^{+0.91} $ & $ 2.07_{-0.47}^{+0.35} $ & $ 1.48_{-0.74}^{+0.71} $ & $ 4.54_{-0.54}^{+0.54} $ \\ 
         &                                               & $s_{+}$              & $ 0.55_{-0.28}^{+0.26} $ & $ 0.19_{-0.10}^{+0.09} $ & $ 3.59_{-1.82}^{+1.75} $ & $ 6.62_{-1.43}^{+0.91} $ & $ 3.51_{-0.80}^{+0.60} $ & $ 1.50_{-0.75}^{+0.70} $ & $ 4.68_{-0.57}^{+0.57} $ \\ 
         & $t_{\rm E} + \theta_{\rm E} + \pivec_{\rm E}$ & $s_{-},\, u_{0}{+}$  & $ 0.27_{-0.09}^{+0.12} $ & $ 0.10_{-0.03}^{+0.05} $ & $ 1.83_{-0.62}^{+0.84} $ & $ 5.26_{-1.01}^{+1.01} $ & $ 1.62_{-0.31}^{+0.33} $ & $ 0.72_{-0.24}^{+0.31} $ & $ 4.69_{-0.60}^{+0.59} $ \\
         &                                               & $s_{-},\, u_{0}{-}$  & $ 0.27_{-0.09}^{+0.12} $ & $ 0.11_{-0.04}^{+0.05} $ & $ 1.99_{-0.69}^{+0.85} $ & $ 5.12_{-0.99}^{+1.00} $ & $ 1.62_{-0.30}^{+0.33} $ & $ 0.74_{-0.24}^{+0.31} $ & $ 4.87_{-0.61}^{+0.62} $ \\
         &                                               & $s_{+},\, u_{0}{+}$  & $ 0.27_{-0.09}^{+0.11} $ & $ 0.10_{-0.04}^{+0.04} $ & $ 1.91_{-0.67}^{+0.83} $ & $ 5.23_{-1.00}^{+0.99} $ & $ 2.76_{-0.51}^{+0.55} $ & $ 0.72_{-0.24}^{+0.30} $ & $ 4.76_{-0.60}^{+0.60} $ \\
         &                                               & $s_{+},\, u_{0}{-}$  & $ 0.27_{-0.09}^{+0.12} $ & $ 0.10_{-0.03}^{+0.05} $ & $ 1.89_{-0.64}^{+0.85} $ & $ 5.14_{-1.00}^{+1.01} $ & $ 2.79_{-0.52}^{+0.57} $ & $ 0.74_{-0.25}^{+0.32} $ & $ 4.81_{-0.62}^{+0.62} $ \\
\hline
KB212294 & $t_{\rm E} + \theta_{\rm E}$                  & C                    & $ 0.11_{-0.06}^{+0.17} $ & $ 0.07_{-0.03}^{+0.10} $ & $ 1.24_{-0.64}^{+1.84} $ & $ 6.86_{-1.06}^{+0.97} $ & $ 0.94_{-0.16}^{+0.16} $ & $ 0.30_{-0.16}^{+0.45} $ & $ 7.63_{-0.77}^{+0.79} $ \\
         &                                               & W                    & $ 0.11_{-0.06}^{+0.17} $ & $ 0.07_{-0.03}^{+0.10} $ & $ 1.23_{-0.64}^{+1.81} $ & $ 6.86_{-1.06}^{+0.98} $ & $ 1.07_{-0.19}^{+0.19} $ & $ 0.31_{-0.16}^{+0.45} $ & $ 7.58_{-0.77}^{+0.79} $ \\
         &                                               & ${\rm R}_{\rm C}$    & $ 0.14_{-0.07}^{+0.20} $ & $ 0.07_{-0.04}^{+0.10} $ & $ 1.27_{-0.65}^{+1.79} $ & $ 6.80_{-1.08}^{+0.97} $ & $ 1.13_{-0.20}^{+0.19} $ & $ 0.38_{-0.19}^{+0.53} $ & $ 7.65_{-0.77}^{+0.79} $ \\
         &                                               & ${\rm R}_{\rm W}$    & $ 0.14_{-0.07}^{+0.20} $ & $ 0.07_{-0.04}^{+0.10} $ & $ 1.26_{-0.65}^{+1.79} $ & $ 6.80_{-1.08}^{+0.97} $ & $ 1.15_{-0.20}^{+0.20} $ & $ 0.38_{-0.19}^{+0.53} $ & $ 7.69_{-0.78}^{+0.81} $ \\
\enddata
\end{deluxetable*}

\begin{figure*}[htb!]
\epsscale{1.00}
\plotone{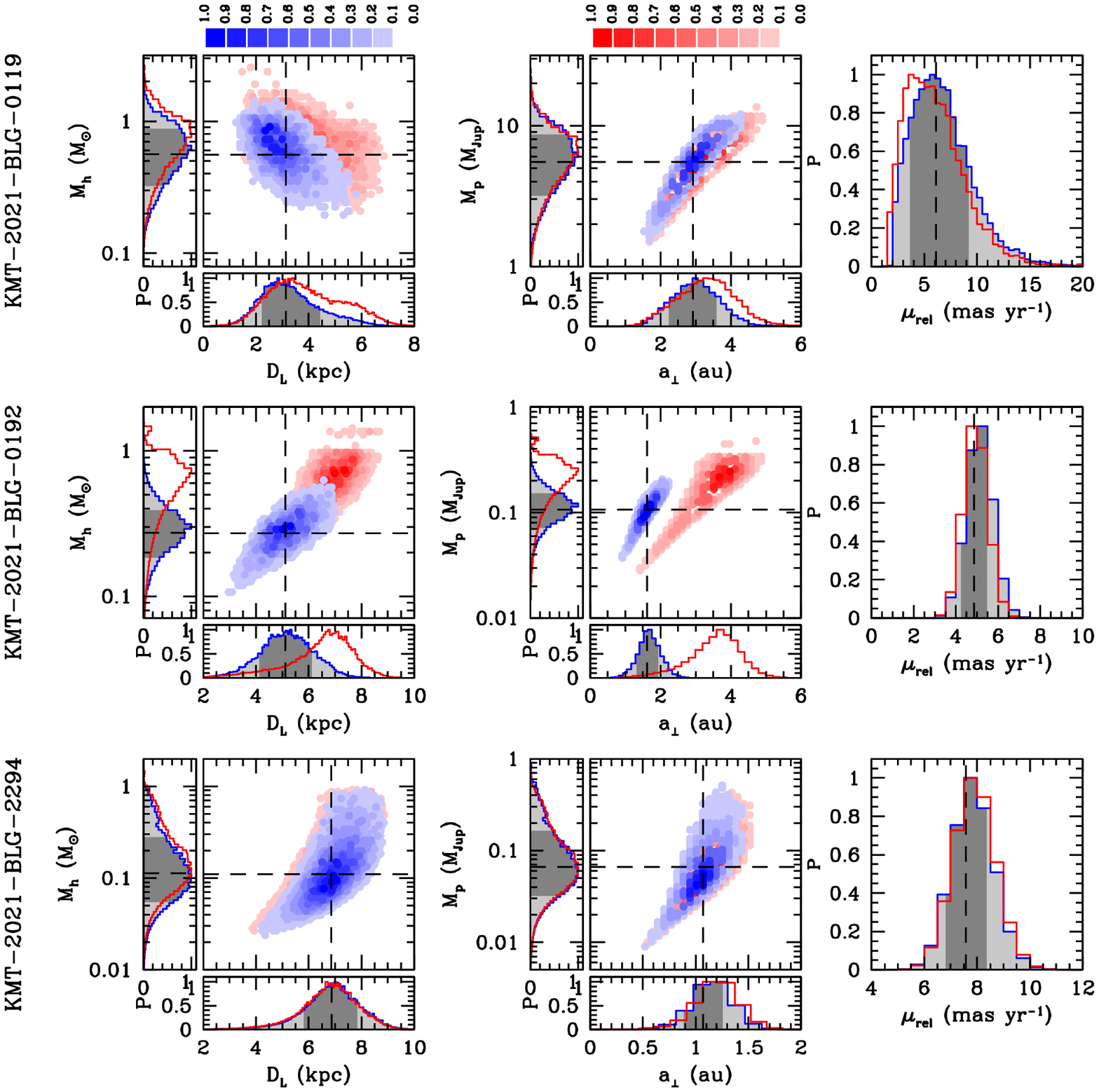}
\caption{The ($D_{L},\, M_{\rm host}$) and ($a_{\perp},\, M_{\rm planet}$) contours with probability 
distributions of the lens properties of each event. The blue contour shows the best-fit case of each 
event. We present an alternate solution as the red contour for comparison. In the histograms, the dark 
gray indicates a $68\%$ confidence interval. The black dashed line indicates the median value of each 
property. For \onenineteen\ (upper panels), we present the Local B family (blue) and Local A family (red). 
For \oneninetwo\ (middle panels), we compare the APRX ($s_{-}$, $u_{0}-$) case (blue; the best fit) with 
the STD $s_{+}$ case (red). For \oneninetwo\ (bottom panels), we present the W case (blue; the best fit) 
and ${\rm R}_{C}$ (red).
\label{fig:lens_properties}}
\end{figure*}

\subsection{Lens Properties of Three events} \label{planet_properties}

In Table \ref{table:lens}, we present the lens properties derived from the Bayesian posteriors for each event. In Figure \ref{fig:lens_properties}, we present the contours of the lens properties with probability distributions of each event. We present the best--fit cases and selected cases for comparison. The plots visualize the possible ranges of the lens properties shown in Table \ref{table:lens}.

For \onenineteen, the lens system consists of a super-Jupiter-mass planet ($M_{\rm planet} \sim 6\, M_{\rm Jup}$) and an early M-type or a K-type dwarf host star ($M_{\rm host} \sim 0.56$ or $\sim 0.69\, M_{\odot}$, for the A and B families of solutions, respectively). The planet orbits the host with a projected separation of $\sim 2.9$ or $\sim 3.2$ au beyond its snow line ($\sim 1.5$ or $\sim 1.9$ au). The planetary system is located at the distance of $\sim 3 - 4$ kpc from us; i.e., half way to the Galactic bulge.

We note that the blend of \onenineteen\, is compatible with the lens posteriors (see Table \ref{table:cmd}). For example, if the lens is an M-dwarf, it would have an absolute magnitude of $M_{I} = 7.2$. Assuming a distance of $3.0$ kpc and that it is behind all of the dust ($A_{I} \sim 0.34$), its observed magnitude would be well matched to the observed blend, which has $I = 19.9$ mag. In the case of the K-dwarf lens ($M_{I} \sim 6.0$ and $D_{\rm L} \sim 4.9$ kpc), the observed magnitude would be also well matched to the observed blend, which has $I = 19.8$ mag. We use the pyDIA reductions to check for an offset between the magnified source and the baseline object, which could show that the blend is not associated with the event. We find $\Delta\thetavec(N,E) = (64, 1.5)\, $ mas. Given that the uncertainties in such measurements are on the order of tens of mas, this measurement does not rule out the possibility that the blend is the lens; i.e., it is not strongly inconsistent with zero. Regardless, because the blend is about $40\%$ of the light, immediate AO followup observations could confirm that the blend is closely aligned to the source. Because the properties of the various solutions are so similar, such observations would not resolve the degeneracy, but they could result in a better characterization of the lens flux.

For \oneninetwo, when including the parallax constraint, the lens system consists of a planet slightly larger than Neptune ($M_{\rm planet} \sim 2\, M_{\rm Nep}$) and an M-dwarf host star ($M_{\rm host} \sim 0.3\, M_{\odot}$). Without the parallax constraint, the values are somewhat larger, but consistent at 1-$\sigma$. The planet is a typical microlensing planet located beyond the snow line. The planetary system is located at $D_{\rm L} \sim 5$ kpc. For completeness, we note that the baseline object appears to be offset from the microlensing event by $\Delta\thetavec(N,E) \sim (-430, 100)\, $ mas so it is not likely to be associated with the event.

For \twotwoninefour, the lens consists of a Neptune-mass planet ($M_{\rm p}\sim 1.2\, M_{\rm Nep}$) orbiting a late M-dwarf host ($M_{\rm h}\sim 0.1\, M_{\odot}$). The system is located in or near the bulge at $D_{\rm L} \sim 6.8\,$ kpc. One interesting point is that the posterior for the host mass significantly overlaps the brown-dwarf regime. This small host mass arises from the short timescale of this event (i.e., $t_{\rm E} \sim$ $7$--$8$ days). In this case, the baseline object appears to be offset from the microlensing event by $\Delta\thetavec(N,E) \sim (160, 410)\, $ mas. This offset implies that the blended light is not due to the lens and that it could be easily resolved from the microlensing target with high-resolution observations. The source itself is reasonably faint ($I = 20.8$ mag), which suggests a contrast ratio of $\Delta K = (2.2, 2.8, 3.5)\, $ mag for a lens mass of $M_{\rm lens} = (0.4, 0.3, 0.2) M_{\odot}$ \citep{BB88, baraffe15}. Given the magnitude of the lens-source relative proper motion ($\sim 7.6 \mathrm{\,mas\, yr}^{-1}$), it should be possible to either measure or place strong upper limits on the lens flux at first light of 30m-class AO systems.

\section{Discussion and Conclusion} \label{sec:discussion}

We have presented three microlensing planets discovered by the KMTNet survey in $2021$: \onenineteen Lb, \oneninetwo Lb, and \twotwoninefour Lb. These planets range in mass from close to a Neptune mass to Super--Jupiter-sized. As is typical of microlensing events, the planet hosts are all likely to be low-mass dwarfs and the systems are $\sim 3$ -- $7\,$ kpc from us. See Table \ref{table:lens}.

Of these three planets, \twotwoninefour\, is the most interesting. First, this event fails the criteria for selection by the AnomalyFinder algorithm \citep{zang21,zang22}. For the AnomalyFinder algorithm, the planet has only $\Delta\chi_0^2 = 37$ for $t_{\rm eff}$ = 0.05, and $\Delta\chi_0^2 = 59$ for $t_{\rm eff} = 0.025$. By contrast, the algorithm has a default threshold of at least $\Delta\chi^2 > 120$. At the same time, the planetary signal is clearly seen by eye in Figure \ref{fig:lc_2294}. Hence, it would be interesting to consider how the algorithm might be modified to detect such signals, although any changes must then be weighed against the potential increase in false positives.

Second, the Bayesian analysis for \twotwoninefour\, suggests that the host is an extremely low-mass M-dwarf. The planet population at this end of the stellar mass function is particularly interesting because of the extreme nature of the hosts. Several studies have suggested that it is more difficult to form giant planets around M-dwarfs via core accretion due to the longer dynamical times \citep{laughlin04,ida05}. For example, the work of \citet{kennedy08} show how giant planet formation varies with stellar mass and suggest that there may be a lower limit on the host mass for giant planets.

\begin{figure*}[htb!]
\epsscale{1.00}
\plotone{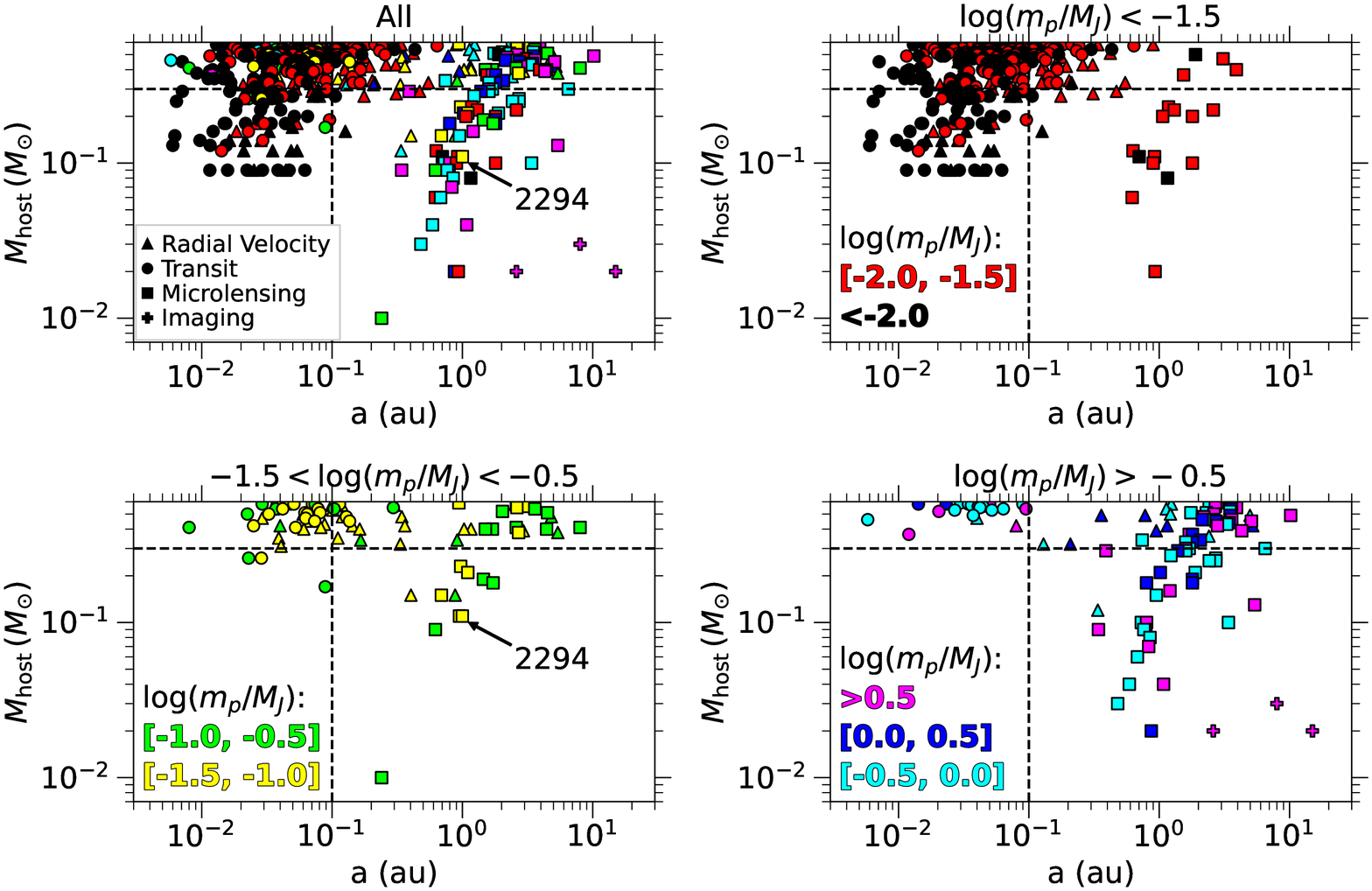}
\caption{Distributions of host mass and system distance for confirmed exoplanets with low-mass host stars. The point shape corresponds to the detection method. The color of each point is set by the (log) mass of the planet (black: $\log(m_p / M_J) < -2$, red: $ -2 < \log(m_p / M_J) < -1.5$, yellow: $ -1.5 < \log(m_p / M_J) < -1$, green: $ -1  < \log(m_p / M_J) < -0.5$, cyan: $ -0.5 < \log(m_p / M_J) < 0$, blue: $ 0 < \log(m_p / M_J) < 0.5$, magenta: $ 0.5 < \log(m_p / M_J)$). For radial velocity planets, $m_p \sin i$ is plotted if $m_p$ is not available, and for microlensing planets $a_{\perp}$ (the projection of the semi-major axis on the sky) is plotted in place of $a$. Reference lines are drawn at $a = 0.1\, $ au and $M_{\rm host} = 0.3\, M_\odot$.  The location of \twotwoninefour\, in this plane is indicated by the arrow. The upper left panel shows all planets together, while the other panels show subsets by planet mass, as indicated in the titles. Data from the NASA Exoplanet Archive, accessed 2022 July 20.}
\label{fig:low-mass-hosts}
\end{figure*}

To place \twotwoninefour Lb in better context with other planets around low-mass host stars, in Figure \ref{fig:low-mass-hosts}, we plot it together with transit, radial velocity, and other microlensing planets from the NASA Exoplanet Archive. This figure shows clear evidence of selection effects, which lead to the appearance of two distinct groups of planets. One group consists of very short-period planets with $a\lesssim 0.1\, $ au whose detections are dominated by the transiting planets, and another group with $a\gtrsim 0.1\,$ au, which is dominated by microlensing planet detections. There is also a trend in the microlensing planets that reflects the fact that $M_{\rm host} \propto a_{\perp}^2$ at fixed $\theta_{\rm E}$.

However, in spite of these selection effects, there is a clear lack of giant planets in the close-in planet population, despite the fact that they should be readily detected. On the other hand, giant planets are abundant in the microlensing sample, which shows planet discoveries at a continuous range of masses. This suggests that there is no particular challenge to forming giant planets around M dwarfs, but there is a challenge for either getting or keeping them in close orbits. 

On the other hand, the majority of the microlensing host masses are derived from Bayesian estimates, although there are a few cases with $M_{\rm L} < 0.3\, M_{\odot}$ for which the lens mass is measured through a combination of $\theta_{\rm E}$ and $\pivec_{\rm E}$ [c.f., OGLE-2017-BLG-1140Lb \citep{calchi18}, OGLE-2017-BLG-1434Lb \citep{udalski18}, OGLE-2018-BLG-0532Lb \citep{ryu20}, and OGLE-2018-BLG-0596Lb \citep{jung19}]. 
In particular, OGLE-2017-BLG-1140L hosts a giant planet that has a well-measured host mass of $0.21\pm0.03\, M_{\odot}$, demonstrating that such planetary systems exist. However, for those events with only Bayesian mass estimates, there is usually a possibility of a more massive host star. As we discussed in Section \ref{sec:lens_properties}, future adaptive optics or other high-resolution imaging of \twotwoninefour\, could confirm that the host mass is indeed $< 0.3 M_{\odot}$. More secure host mass measurements or limits for the microlensing planet population would allow for a study of how the planet distribution varies with host mass, which could then be linked back to planet formation theory \citep[e.g.,][]{kennedy08} and compared to radial velocity studies \citep[e.g.,][]{bonfils13}.

\mbox{}

This research has made use of the KMTNet system operated by the Korea Astronomy and Space Science Institute (KASI), and the data were obtained at three host sites of CTIO in Chile, SAAO in South Africa, and SSO in Australia.  
I.-G.S., S.-J.C., and J.C.Y. acknowledge support from N.S.F Grant No. AST--2108414.
Work by C.H. was supported by grants of the National Research Foundation of Korea (2017R1A4A1015178 and 2019R1A2C2085965).
Y.S. acknowledges support from BSF Grant No. 2020740.
The MOA project is supported by JSPS KAKENHI Grant Number JSPS24253004, JSPS26247023, JSPS23340064, JSPS15H00781, JP16H06287, and JP17H02871.
The computations in this paper were conducted on the Smithsonian High Performance Cluster (SI/HPC), Smithsonian Institution (\url{https://doi.org/10.25572/SIHPC}).
This research has made use of the NASA Exoplanet Archive, which is operated by the California Institute of Technology, under contract with the National Aeronautics and Space Administration under the Exoplanet Exploration Program.


\newpage

\end{document}